\documentclass[twocolumn,trackchanges]{aastex7}
\usepackage{lmodern}
\usepackage{ae,aecompl,multirow}
\usepackage{threeparttable}
\usepackage{booktabs}
\usepackage{url}
\usepackage{amsmath}
\usepackage{graphics,comment}
\usepackage{textcomp} 
\usepackage{hyperref} 

\usepackage{booktabs}  
\DeclareUnicodeCharacter{2212}{\textminus}
\begin{document}

\title{Spectroscopic Reverberation Mapping for SARM: The Case of Mrk 1048 and Mrk 618}

\correspondingauthor{Shivangi Pandey}
\author[0000-0002-4684-3404]{Shivangi Pandey}
\affiliation{Aryabhatta Research Institute of Observational Sciences, Nainital\textendash263001, Uttarakhand, India}
\affiliation{Department of Applied Physics/Physics, Mahatma Jyotiba Phule Rohilkhand University, Bareilly\textendash243006, India}
\email[show]{Shivangipandey@aries.res.in}

\author[0000-0002-8377-9667]{Suvendu Rakshit}
\affiliation{Aryabhatta Research Institute of Observational Sciences, Nainital\textendash263001, Uttarakhand, India}
\email[show]{suvenduat@gmail.com}

\author[0000-0002-4024-956X]{S. Muneer}
\affiliation{Indian Institute of Astrophysics, Bengaluru 560034, India}
\email{muneers@iiap.res.in}

\author{Jincen Jose}
\affiliation{Aryabhatta Research Institute of Observational Sciences, Nainital\textendash263001, Uttarakhand, India}
\affiliation{Center for Basic Sciences, Pt. Ravishankar Shukla University, Raipur, Chhattisgarh\textendash492010, India}
\email{jincenjose@gmail.com}

\author{Ashutosh Tomar}
\affiliation{Aryabhatta Research Institute of Observational Sciences, Nainital\textendash263001, Uttarakhand, India}
\affiliation{Department of Physics, Indian Institute of Technology Roorkee, Roorkee 247667, Uttarakhand, India}
\email{1101ashutosh@gmail.com}

\author[0000-0001-5841-9179]{Yan-Rong Li}
\affiliation{Institute of High Energy Physics, Chinese Academy of Sciences, Beijing 100049, China}
\email{liyanrong@ihep.ac.cn}

\author[0000-0001-9449-9268]{Jian-Min Wang}
\affiliation{Institute of High Energy Physics, Chinese Academy of Sciences, Beijing 100049, China}
\email{wangjm@ihep.ac.cn}

\author[0000-0002-4998-1861]{C. S. Stalin}
\affiliation{Indian Institute of Astrophysics, Block II, Koramangala, Bangalore\textendash560034, India}
\email{cs.stalin@gmail.com}

\author[0000-0002-8055-5465]{Jong-Hak Woo}
\affiliation{Department of Physics \& Astronomy, Seoul National University, Seoul\textendash08826, Republic of Korea}
\email{woo@astro.snu.ac.kr}

\author[0000-0003-4759-6051]{Romain G. Petrov}
\affiliation{Lagrange Laboratory, Université Côte d'Azur, CNRS, Observatoire de la Côte d'Azur, Nice, France}
\email{romain.petrov@unice.fr}

\author[0000-0001-6009-1803]{James Leftley}
\affiliation{School of Physics and Astronomy, University of Southampton, Southampton SO17 1BJ, United Kingdom}
\email{J.Leftley@soton.ac.uk}

\author[0000-0002-6353-1111]{Sebastian F. Hönig}
\affiliation{School of Physics and Astronomy, University of Southampton, Southampton SO17 1BJ, United Kingdom}
\email{s.hoenig@soton.ac.uk}

\author[0000-0001-9957-6349]{Amit Kumar Mandal}
\affiliation{Center for Theoretical Physics of the Polish Academy of Sciences, Al. Lotnik$\Acute{o}$w 32/46, 02\textendash668 Warsaw, Poland}
\email{amitastroam@gmail.com}

\author{Tushar Ubarhande}
\affiliation{Department of Applied Optics and Photonics, University of Calcutta, Technology Campus, JD\textendash2, Sector 3, Bidhannagar, Kolkata \textendash700106 West Bengal}
\affiliation{Aryabhatta Research Institute of Observational Sciences, Nainital\textendash263001, Uttarakhand, India}
\email{tushar.g.ubarhande@gmail.com}

\author[0000-0002-2052-6400]{Shu Wang}
\affiliation{Department of Physics \& Astronomy, Seoul National University, Seoul\textendash08826, Republic of Korea}
\email{wangshu100002@gmail.com}

\author[0000-0002-1207-0909]{Michael Brotherton}
\affiliation{Department of Physics and Astronomy, University of Wyoming, Laramie, WY 82071, USA}
\email{MBrother@uwyo.edu}

\author{Archana Gupta}
\affiliation{Department of Applied Physics/Physics, Mahatma Jyotiba Phule Rohilkhand University, Bareilly\textendash243006, India}
\email{archana.gupta@mjpru.ac.in}
\begin{abstract}
Robust extragalactic distance measurements are crucial for resolving the persistent discrepancy in the value of the Hubble constant (\(H_0\)). Active Galactic Nuclei (AGNs), through their compact and variable broad-line regions (BLRs), enable the determination of geometric distances when reverberation mapping (RM) is combined with spectroastrometry(SA). We report results from a spectroscopic RM campaign (October 2022 to March 2023) targeting two narrow-line Seyfert 1 galaxies, Mrk 1048 and Mrk 618, using 3.6-m DOT and 2-m HCT. High-cadence spectro-photometric monitoring was carried out using onboard instruments such as ADFOSC, HFOSC, and TANSPEC, resulting in well-sampled continuum and emission line light curves. The observed fractional variability ($F_{\mathrm{var}}$) ranged from 4\% to 14\% across the $g$-band, H$\beta$, and H$\alpha$ light curves. The time lags were measured using the interpolated cross-correlation function (ICCF), PyI$^{2}$CCF, and \textsc{JAVELIN} methods. In the rest frame, the ICCF analysis yields H$\beta$ lags of $10.5^{+2.6}_{-4.2}$ days for Mrk 1048 and $10.2^{+3.4}_{-2.9}$ days for Mrk 618, while the corresponding H$\alpha$ lags are $18.7^{+5.3}_{-5.4}$ and $14.4^{+4.6}_{-10.5}$ days, respectively. The emission-line widths, measured from the rms spectra using $\sigma_{\mathrm{line}}$, give virial black hole mass estimates of $6.3^{+2.0}_{-2.1} \times 10^7\,M_\odot$ for Mrk 1048 and $1.2^{+0.4}_{-0.6} \times 10^7\,M_\odot$ for Mrk 618. These results will serve as a basis for absolute geometric distance calibration when combined with VLTI/GRAVITY spectro-astrometric measurements, thereby contributing to the development of AGNs as standardizable cosmological probes.
\end{abstract}
\keywords{Reverberation Mapping, Active galactic nuclei, Black holes, Optical telescopes, Distance measure}
\section{Introduction} \label{sec:intro}
Active Galactic Nuclei (AGNs) are among the most luminous and enduring objects in the universe, powered by accretion onto supermassive black holes (SMBHs) \citep{Rees1984} with masses exceeding $10^6 M_\odot$ \citep{Woo2002}. Their characteristic broad emission lines and strong continuum variability have long served as vital probes of the central regions of AGNs. A particularly promising geometric approach to studying the broad-line region (BLR) structure \citep{Elvis_2002, Rakshit2015} and measuring AGN distances is the combined use of spectroastrometry and reverberation mapping (SARM) that was first introduced in \citet{Wang2020}. Spectroastrometry (SA) enables sub-diffraction-limit angular measurements by tracing wavelength-dependent photocenter displacements \citep{Rakshit2015}, while reverberation mapping (RM) provides radial BLR sizes based on the time delay between variations in the ionizing continuum and the corresponding response in emission lines. Together, these techniques enable not only to constrain the geometry and kinematics and measure the black hole masses, but also to estimate the geometric distance independent of the traditional cosmic distance ladder. Applications to a few objects yielded $H_0$ values consistent with standard cosmology \citep{Gravity2021, Li2024, Li2025}. Further theoretical refinements, including modeling of BLR emissivity and responsivity, have improved the accuracy to within $\sim$10--30\% \citep{LiandWang2023, Li2025}, although the precision remains constrained by current interferometric capabilities.

RM \citep{Bahcall1972, Blandford1982, Peterson_1993}, which is a traditional method for studying the central engine of AGNs, uses the time delay between variations in the ionizing continuum (from the accretion disk) and the broad emission lines (from the BLR) to measure the size of the BLR. Assuming the BLR gas is virialized, the SMBH mass can be calculated using the virial equation:

\begin{align}\label{eq:virial eqn}
M_{\mathrm{BH}} =  \dfrac{f \times R_{\mathrm{BLR}} (\Delta V)^{2}}{G}
\end{align}

where $R_{\mathrm{BLR}}$ is the BLR radius (from time lag), $\Delta V$ is the line width (FWHM or $\sigma_{\mathrm{line}}$), and $f$ is the virial factor inferred from BLR geometry and inclination. RM has been applied to over a hundred AGNs, successfully calibrating the $R_{\mathrm{BLR}} - L_{5100}$ relation \citep{Kaspi2000, bentz_low-luminosity_2013, Cho2023, Woo2024, Sobrino2025}, and yielding insights into black hole mass scaling relations, AGN structure, and accretion physics \citep[e.g.,][]{shen_catalog_2011, Du2015, Du2016, Pei2017, Grier2017, Rakshit2019, Cackett2021, U2022, Cho2023, Shen2024, Woo2024, Wang_2024, Sobrino2025}. The velocity-resolved delay map and dynamical modelling of RM data showed evidence for Keplerian rotation of the BLR clouds and a disc-like BLR in many AGNs \citep[e.g.,][]{Pancoast2014, Li2018, Wang_2025}. However, these are limited by the requirements of better and higher cadence data.

Despite its success, RM is fundamentally limited by its inability to resolve full spatial structures, as it probes only line-of-sight velocities. This constraint has been significantly alleviated by advances in optical/IR interferometry. In particular, the GRAVITY instrument on the VLTI has spatially resolved BLRs in nearby AGNs with $\sim$10\,$\mu$as precision, observing sources like 3C~273 and NGC~3783 \citep{Gravity2018, Gravity2020, Gravity2021}. GRAVITY has also revealed a strong correlation between hot dust sizes and RM-based BLR radii, offering an alternative path for SMBH mass estimation with fewer observational demands \citep{Gravity2024}. Combined with RM, this enables a complementary spatial–temporal view of BLR geometry. 

To fully exploit SARM for distance measurements and SMBH mass estimation, high-quality RM data remains essential. While GRAVITY/VLTI provides angular sizes for nearby AGNs, accurate BLR linear sizes from RM are needed to derive angular diameter distances. This requires long-term, high-cadence spectroscopy of AGNs with strong, variable broad lines. Motivated by this, we initiated an RM campaign targeting AGNs observable with GRAVITY/VLTI, aiming to measure time lags between the ionizing continuum and broad-line variations. These lags yield estimates of the BLR radii and virial mass for future SARM studies. This paper presents the campaign’s initial results, variability analysis, lag measurements, and black hole mass estimates. Section~\ref{sec:Observation} details the target selection and observations; Section~\ref{sec:analysis} describes data processing and light curve analysis; Section~\ref{sec:timelag analysis} presents lag measurements via ICCF and \textsc{JAVELIN}; Section~\ref{sec:Black hole mass measurement} covers mass estimation methods; Section~\ref{sec:Discussion} offers comparisons and implications; and Section~\ref{sec:conclusion} summarizes our findings.
\section{Target selection and Observations}
\label{sec:Observation}
To assemble a sample suitable for SARM observations from both hemispheres, we began with the catalog of \citet{Wang2020}, which lists 30 low-redshift AGNs ($z < 0.08$) with $K < 11.5$ and expected BLR angular sizes $\gtrsim$20$\mu$as. These characteristics make them promising candidates for distance measurements using strong Brackett~$\gamma$ or Paschen~$\alpha$ emission lines accessible to GRAVITY. Among these 30 AGNs, only seven have declinations higher than $-15^\circ$ and $K$-band magnitudes brighter than 11, rendering them accessible to ground-based observatories in both hemispheres. Initially, we planned to monitor all seven sources using the 3.6-m Devasthal Optical Telescope \citep[DOT;][]{Kumar2018}, ARIES, Nainital and 2-m Himalayan Chandra Telescope (HCT) at the Indian Astronomical Observatory (IAO), Hanle, India, as part of this campaign. Over 5-6 months, i.e., October 2022 to March 2023, four AGNs were successfully observed. However, data quality was affected by external factors such as seasonal gaps, weather conditions, variability constraints, and cadence issues. Ultimately, only two out of four sources, Mrk 1048 and Mrk 618, exhibited well-sampled light curves with better cadence spectroscopic monitoring and strong correlation properties. The Table \ref{tab:mrk_obs} depicts the different properties for Mrk 1048 and Mrk 618, including redshift, luminosity distance, magnitude in $V$-band, corrected for extinction value, total number of epochs of spectro-photometric observation, the cadence in both, the period of observations, and telescopes.

\begin{table*}
    \centering
    \caption{Observational details of Mrk 1048 and Mrk 618}
    \begin{tabular}{l c c c c cc cc cc c}
        \toprule
        Name & $z$ & $D_L$ & $m_V$ & \multicolumn{2}{c}{$N_{\rm obs}$} & \multicolumn{2}{c}{$\Delta t_{\rm med}$} & \multicolumn{2}{c}{MJD} & Telescopes \\
        &  & (Mpc) & (mag) &Spec & Phot & Spec & Phot & Spec & Phot & \\
        (1) & (2) & (3) & (4) & (5) & (6) & (7) & (8) & (9) & (10) & (11) \\
        \midrule
        Mrk 1048 & 0.0426 & 191.4 & 14.02 & 25 & 125 & 7 & 1-3 & 59856.79-59998.81 & 59796.33-60001.07  & DOT, HCT \\
        Mrk 618  & 0.0355 & 154.9 & 14.10 & 25 & 211 & 7 & 1-3 & 59856.87-60014.81 & 59796.27-60030.60 & DOT, HCT \\
        \bottomrule
    \end{tabular}
    \label{tab:mrk_obs}
 \begin{flushleft}
        \textbf{Note:} Column (1): Object name. Column (2): redshift from NASA/IPAC Extragalactic Database (NED). Column (3): luminosity distance derived from redshift. Column (4): $V$-band magnitude. Columns (5)–(6): number of observation epochs (spectroscopic and photometric, including ZTF and ASAS-SN). Columns (7)–(8): median sampling interval (spectroscopic and photometric). Columns (9)–(10): Duration of observations (spectroscopic and photometric). Column (11): telescopes used.
\end{flushleft}
\end{table*}

Weekly cadence spectroscopic and photometric observations of both sources were carried out using optical and near-infrared (NIR) spectrographs mounted on the 3.6-m DOT and the 2-m HCT. Optical spectro-photometric data with DOT were obtained using the ARIES Devasthal Faint Object Spectrograph and Camera (ADFOSC) \citep{omar_first-light_2019}. ADFOSC features a 4K$\times$4K deep-depletion CCD camera, yielding a pixel scale of 0.2$^{\prime\prime}$/pixel with 2$\times$2 binning \citep{Dimple2022}. However, due to technical reasons (limited ports to mount the instrument at DOT), the optical spectrograph ADFOSC, which covers both the H$\beta$ and H$\alpha$ lines, is available for $\sim$2 months in each observation cycle. Therefore, we have also used the TIFR-ARIES Near Infrared Spectrometer (TANSPEC) \citep{Sharma2022} (mounted at DOT alternatively with ADFOSC), which covers the optical H$\alpha$ line ($\mu$m) along with other infrared emission lines. TANSPEC is equipped with two Teledyne HgCdTe Astronomical Wide Area Infrared Imager (HAWAII) detectors: an H1RG (1024$\times$1024 pixels) for imaging and slit viewing, and an H2RG (2048$\times$2048 pixels) for spectroscopy. The instrument offers a 1$\times$1 arcmin$^2$ field of view and covers a wavelength range of 0.55–2.5 $\mu$m, split into 10 spectral orders.

Optical observations with the HCT utilized the Himalayan Faint Object Spectrograph Camera (HFOSC)\footnote{\url{https://www.iiap.res.in/centers/iao/facilities/hct/}}, a versatile instrument designed for low- and medium-resolution grism spectroscopy. The detector comprises a SITe ST-002 2K$\times$4K pixel CCD, with the central 2K$\times$2K region used for imaging. This setup provides a field of view (FOV) of approximately 10$^{\prime} \times$10$^{\prime}$ and a pixel scale of 0.296$^{\prime\prime}$/pixel.
\begin{figure*}[t]
\includegraphics[scale=0.40]{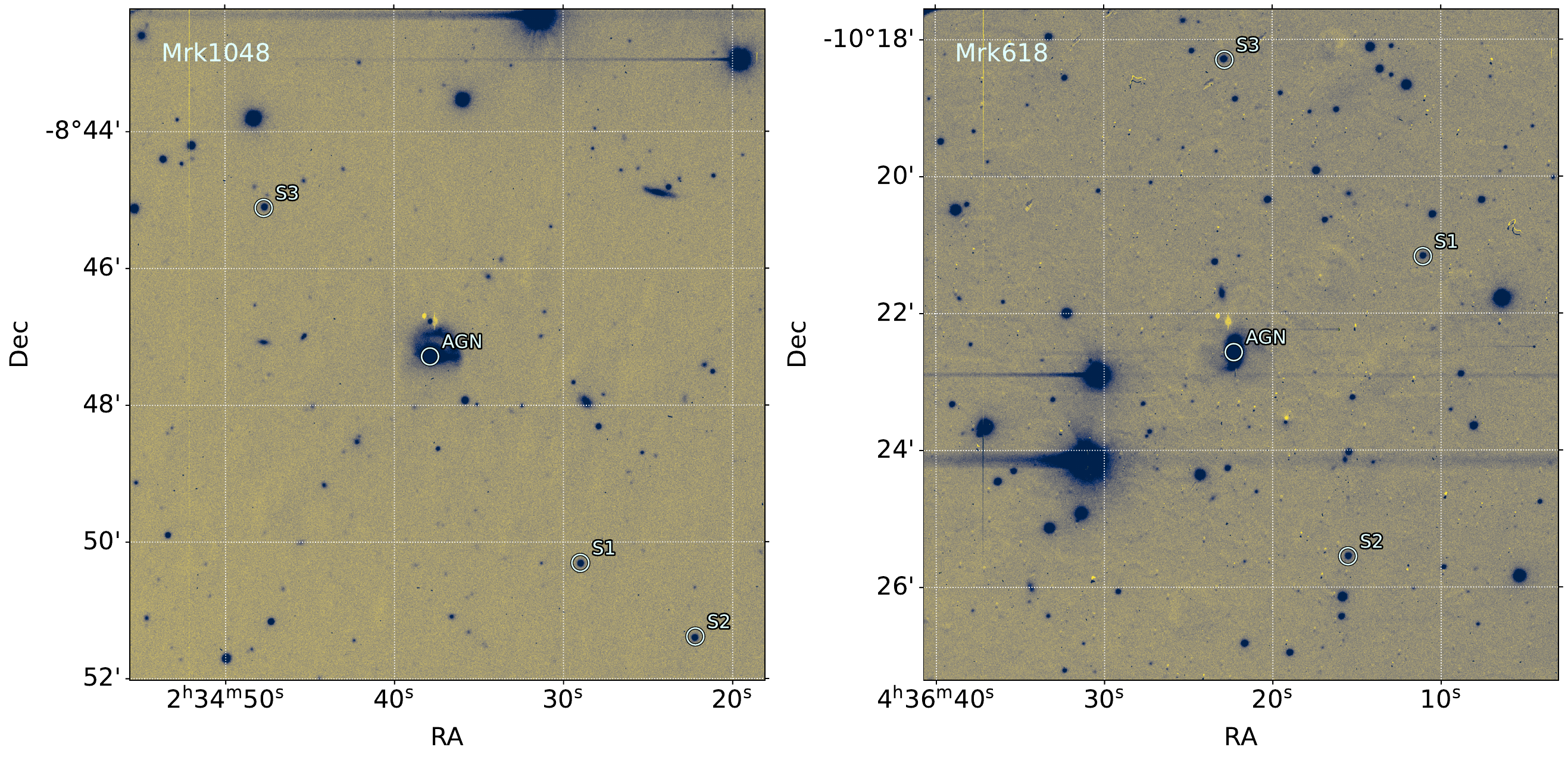}
\caption{$V$-band image of Mrk 1048 and Mrk 618 observed from the HFOSC/HCT with a field of view of 10$\times$10$^{\prime}$. The sources are marked, and the nearby comparison stars are shown.}
\label{fig:V_frame}
\end{figure*}
\subsection{Photometry} \label{sec:Photometry}
Photometric observations were conducted using broadband filters, specifically the SDSS $r$-band (623 nm) from ADFOSC, the $V$-band (550 nm) from HFOSC and the $R$-band (612 nm) from TANSPEC. For each target, three broadband photometric frames were acquired with exposure times ranging from 30 to 60 seconds, immediately preceding the spectroscopic observations. Fig.~\ref{fig:V_frame} presents the $V$-band images of Mrk 1048 and Mrk 618, with the central AGN marked. Images obtained from each observation night were initially aligned using the Astroalign Python package \citep{Astroalign}. The pre-processing of photometric frames followed standard procedures, including bias subtraction, flat-field correction, and cosmic ray removal. Aperture photometry was carried out using SEP, a Python-based wrapper for the Source Extractor package \citep{Barbary2016}. Differential photometry was performed by selecting 3–5 nearby reference stars, as shown in Fig.~\ref{fig:V_frame}. The photometric aperture was set to 2.5 times the average full width at half maximum (FWHM) of the selected comparison stars, where the FWHM was determined by fitting a Gaussian profile to the data. The local sky background was estimated within an annular region extending from 4 to 5 times the FWHM. The differential magnitude of the source was then calculated relative to the comparison stars in the same field of view. Finally, a photometric zero point was applied to convert instrumental magnitudes into calibrated broadband magnitudes.

In addition to our observations, we incorporated archival $g$-band photometric data from two public time-domain surveys: the All-Sky Automated Survey for Supernovae (ASAS-SN) \citep{Kochanek2017} and the Zwicky Transient Facility (ZTF) \citep{Graham2019}. To ensure photometric quality, we selected only high-quality measurements by applying a flag condition of bad catflags ask $ = 0$, which excludes data points marked with any known issues such as saturation, blending, or poor centroiding. This effectively removes outliers and spurious detections from the ZTF light curves. These datasets span the period from May 2022 to March 2023. Given the higher cadence of the ASAS-SN $g$-band data (centered at 4747{\AA}), we intercalibrated all photometric measurements from other bands to the ASAS-SN scale using the {\tt PyCALI} software \citep{Pycali2014}. This intercalibration significantly enhanced the temporal sampling of the light curves, as summarised in Table~\ref{tab:mrk_obs}. For the subsequent analysis, we adopt the intercalibrated $g$-band light curve as the primary continuum-driving signal.
\subsection{Spectroscopy} \label{sec:Spectroscopy}
Spectroscopic observations were conducted using three instruments: ADFOSC and TANSPEC onboard the 3.6-m DOT, and HFOSC onboard the 2-m HCT, each configured to obtain high-quality spectral data. Below, we detail the setup and reduction procedures used for each instrument.

1) ADFOSC (mounted at 3.6-m DOT) spectroscopic observations were performed using a 1.2$^{\prime\prime}$-wide and 8$^{\prime\prime}$-long slit in combination with a 132R-600 gr/mm grism, covering the wavelength range 3500–7000 {\AA}, and centered at 4880{\AA}. Each spectroscopic frame had an exposure time of 600 seconds. Bias and flat-field frames were also acquired throughout the night for standard calibration. Seeing conditions during the observations ranged between 0.5$^{\prime\prime}$ and 1.5$^{\prime\prime}$. The instrumental resolution was determined to be 7{\AA} (corresponding to 312 km s$^{-1}$), measured by modeling the emission lines in a combined Hg-Ar-Ne arc lamp frame taken with the same configuration as the science exposures.

2) HFOSC (mounted at 2-m HCT) spectroscopic data were obtained using Grism 7, which provides a spectral resolution of $R = \lambda/\Delta\lambda \sim 1320$ and covers a wavelength range of 3800–6840{\AA}. Observations employed a 1.15$^{\prime\prime}$-wide and 11$^{\prime}$-long slit. Wavelength calibration was carried out using Fe-Ar and Fe-Ne hollow cathode lamps, taken immediately before and after the science frames. The observations were conducted under good photometric conditions, with an average full width at half maximum (FWHM) seeing of approximately 1.6$^{\prime\prime}$. Bias and flat-field calibration frames were collected at the beginning and end of each night. The resulting spectral resolution achieved was 8{\AA}.

\begin{figure*}[t]
    \centering
    \includegraphics[scale=1,width=6.2in,height=3.5in]{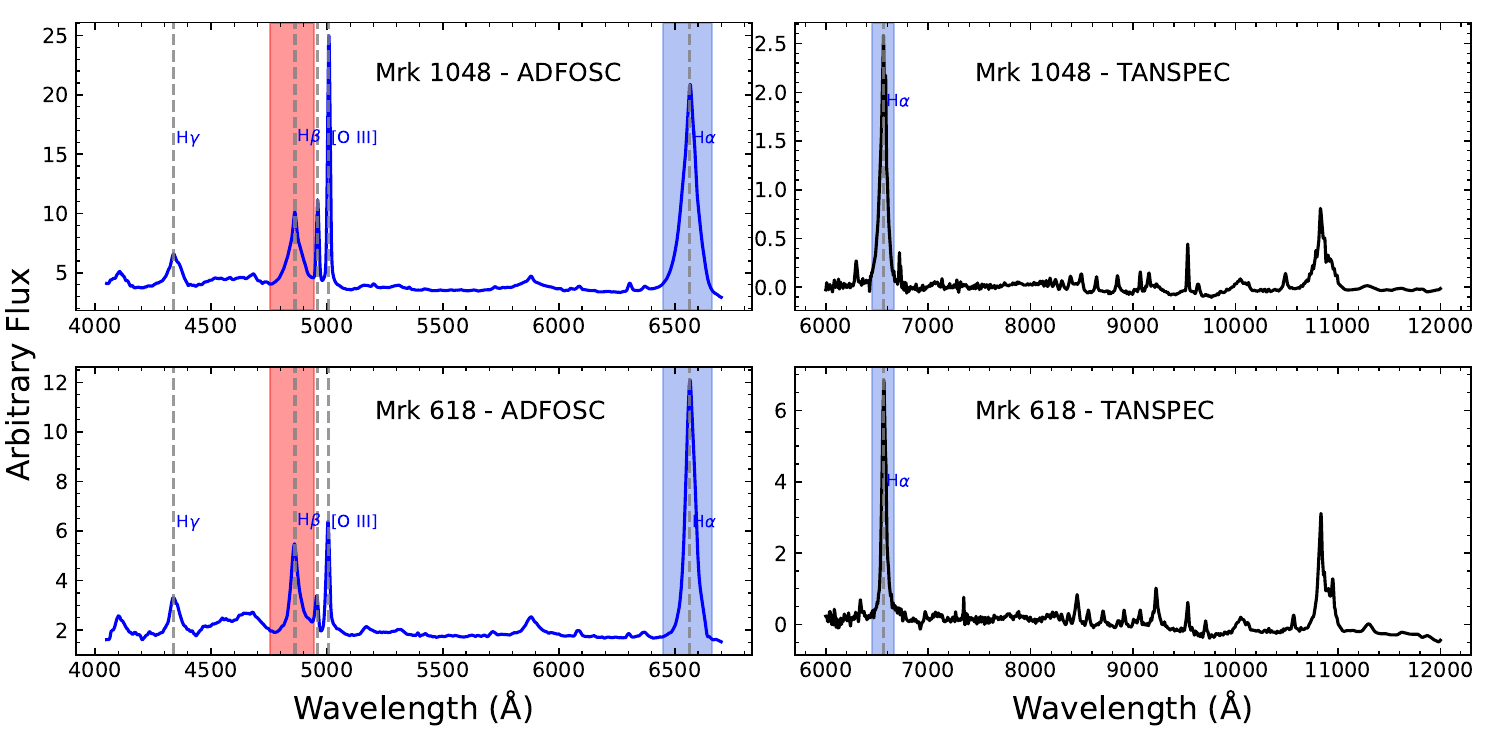}
    \caption{The composite spectra from ADFOSC (left) and TANSPEC (right) of Mrk 1048 (top) and Mrk 618 (bottom) are shown. The emission line regions, such as H$\gamma$ and H$\beta$, H$\alpha$, with narrow emission lines [\ion{O}{3}] are highlighted.}
    \label{fig:mean_spectrum}
\end{figure*}

Spectroscopic data reduction was performed using IRAF \citep{Tody1986, Tody1993, IRAF}, following standard procedures including bias subtraction, flat-fielding, and cosmic ray removal using the L.A. Cosmic algorithm \citep{dokkum_cosmicray_2001}. Flat-field correction was applied using a tungsten-LED lamp for ADFOSC and a halogen lamp for HFOSC, with both lamps normalized before division by the science frames. Spectral extraction was carried out using the 'apall' task in IRAF, with an aperture size of 7–8$^{\prime\prime}$ set for both the target source and comparison stars. Wavelength calibration for ADFOSC spectra was performed using Hg-Ar and Neon arc lamps, whereas Fe-Ar and Fe-Ne lamps were used for HFOSC data. All calibration lamps were observed in the same instrumental configuration as the respective science exposures. For ADFOSC, the calibration lamp frames were combined using the 'imcombine' task in IRAF. The resulting combined calibration spectrum was then used to derive the wavelength solution, which was subsequently applied to both the science and reference star spectra. Flux calibration was achieved using a spectrophotometric standard star, from which a sensitivity function was derived and applied. Fig.~\ref{fig:mean_spectrum} displays the mean spectrum obtained from all ADFOSC observations, with prominent emission line regions indicated. The upper and lower left panels correspond to Mrk 1048 and Mrk 618, respectively.

3)The TANSPEC (mounted at 3.6-m DOT) spectrograph has two operational modes, capturing spectral data on a 2K $\times$ 2K H2RG array. We have carried out spectroscopic observations in the cross-dispersed (XD) mode, which uses a combination of a grating and two prisms that are employed to pack all spectral orders onto the H2RG detector, achieving a resolution of $R \sim 1500$ for a 1$^{\prime\prime}$ slit width. Standard observational procedures were followed: each target was nodded along the slit at two positions, with multiple exposures taken at each nod to enhance the signal-to-noise ratio (SNR). Exposure times were limited to three minutes per frame to facilitate effective cancellation of telluric emission lines through frame differencing at alternating nod positions. Telluric correction was performed using a nearby A0V-type standard star. Additionally, argon and neon arc lamps were used for wavelength calibration, and tungsten lamps were employed for flat-fielding. Calibration frames were acquired for each target to ensure accurate and precise spectral calibration. Data reduction was carried out using the \texttt{pyTANSPEC} pipeline \citep{Ghosh2023}\footnote{\url{https://github.com/astrosupriyo/pyTANSPEC}}, a dedicated tool designed for reducing TANSPEC cross-dispersed (XD) mode spectra. The extracted spectra were corrected for telluric absorption features and subsequently normalized. Continuum-normalized spectra from different spectral orders were combined to construct a final composite spectrum for each source. Flux calibration of the TANSPEC spectra was performed using a telluric standard star and photometric observations taken around the same time, and the spectral slope was adjusted based on data from the HCT or ADFOSC instrument on DOT. A nearby epoch spectrum from HCT or DOT was used to measure the slope of the source continuum, which was then applied to the wavelength-calibrated TANSPEC spectrum. After this correction, the spectrum was rescaled to match the corresponding photometric flux, ensuring consistency between the spectral and photometric data. Fig.~\ref{fig:mean_spectrum} shows the composite mean spectra of Mrk 1048 and Mrk 618 in the upper and lower right panels, respectively. The H$\alpha$ emission line and several Paschen series lines are prominently visible. However, in Fig.~\ref{fig:mean_spectrum} only a small portion of the spectrum is visible to highlight the H$\alpha$ emission line. The spectra are comparatively noisier due to residual telluric features and gaps in atmospheric transmission. 
\section{Analysis}\label{sec:analysis}
\subsection{Spectral Decomposition}
\label{sec:Spec Decomposition}
Accurate RM measurements require precise estimation of the intrinsic variability of AGN, necessitating correction for extrinsic factors such as changes in observing conditions. Failure to account for such effects can result in the misattribution of variability to intrinsic AGN emission. To mitigate this, we employed the {\tt mapspec} Python package\footnote{\url{https://github.com/mmfausnaugh/mapspec/}} \citep{MAPSPEC}, which implements a Bayesian framework based on the method of \citet{van_Groningen1992}. It applies an empirical template to correct time-series spectra for variations in wavelength calibration, attenuation, and spectral resolution. For spectral calibration, the [\ion{O}{3}]$\lambda$5007 emission line was used as a non-variable reference feature to rescale spectra and correct for redshift shifts at each epoch for both sources. The [\ion{O}{3}] extraction windows were defined as [4984, 5025]{\AA} for Mrk 1048 and [4990, 5027]{\AA} for Mrk 618, to account for source-specific linewidth differences. The adjacent continuum windows were set as [4980, 4990]{\AA} and [5027, 5037]{\AA} for Mrk 1048, and [4974, 4984]{\AA} and [5025, 5035]{\AA} for Mrk 618.

The {\tt mapspec} package standardized the [\ion{O}{3}] profiles across all epochs by correcting for wavelength shifts, flux scaling, and line broadening using a Gauss-Hermite kernel \citep{Fausnaugh2017}. The reference epoch was selected based on the spectrum exhibiting the broadest [\ion{O}{3}] profile and using HCT spectral [\ion{O}{3}] flux having a large number of data points comparatively, typically corresponding to data acquired under poor seeing conditions or affected by slit losses. Calibration uncertainties were measured using a Markov Chain Monte Carlo (MCMC) approach. The resulting flux scaling uncertainties were combined in quadrature with the measurement errors of the H$\beta$ flux, ensuring accurate propagation of calibration errors into the final light curve.
\begin{figure}[htbp]
    \centering
    \includegraphics[width=0.95\linewidth]{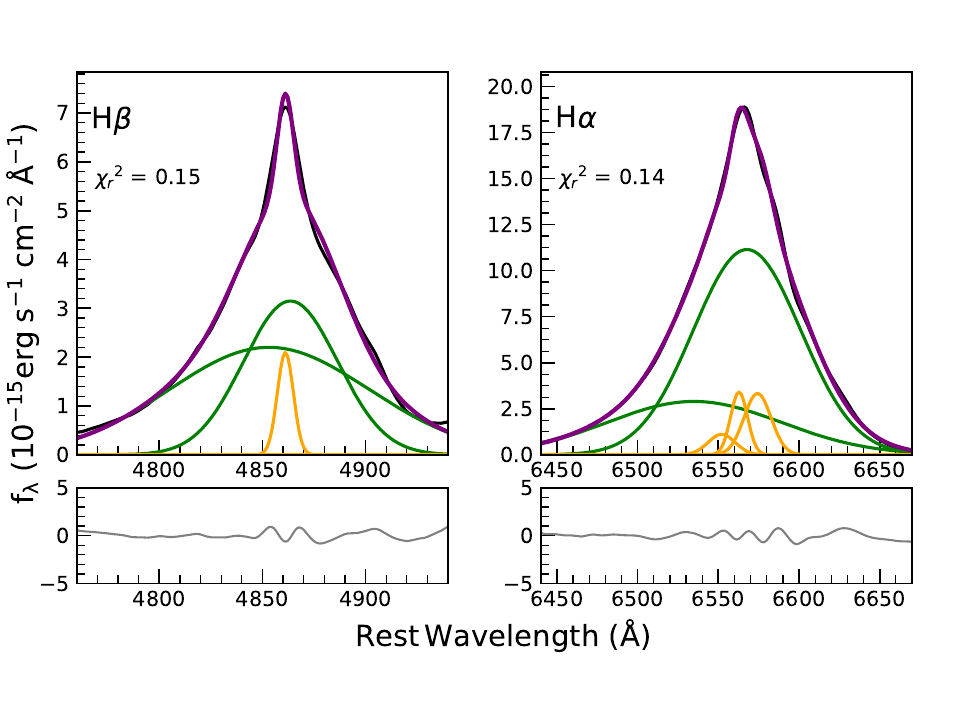}
    \caption{L-R: The H$\beta$ and H$\alpha$ emission line plots are shown along with their residual after fitting for Mrk 1048. The broad component fitting with a double Gaussian is in green, whereas the narrow component fitting is shown in orange. The total line model is overplotted on the original continuum-subtracted spectrum. The H$\alpha$ emission line fiting have decomposed the [\ion{N}{2}]$\lambda$6549, and [\ion{N}{2}]$\lambda$6585 and narrow H$\alpha$ component.}
    \label{fig:model}
\end{figure}
The publicly available multi-component spectral fitting code PyQSOFit, developed by \citet{GuoPyqsofit} and \citet{guo_legolasonpyqsofit_2023}, was employed for spectral decomposition and line fitting. A comprehensive description of the code and its applications can be found in \citet{guo_constraining_2019}, \citet{shen_sloan_2019}, and \citet{Rakshit2020ApJS}. Each AGN spectrum was first corrected for Galactic extinction using the reddening map of \citet{Schlegel1998} and the Milky Way extinction law of \citet{Fitzpatrick1999}, adopting $R_V = 3.1$. The spectra were then de-redshifted using the redshifts listed in Table~\ref{tab:mrk_obs}. The underlying continuum was modeled with a power-law fit over selected line-free regions of the spectrum, specifically: [4200, 4260], [4435, 4640], [5100, 5535], [6005, 6035], and [6110, 6250]{\AA}. Additionally, FeII emission was modeled using templates from \citet{Boroson1992}. The signal-to-noise ratio (S/N) in continuum regions around 5100{\AA} was typically in the range of 20–30 for both sources. Following continuum subtraction, detailed multi-Gaussian modeling was performed in the H$\beta$ and H$\alpha$ regions, as illustrated in Fig.~\ref{fig:model}. The narrow components of H$\beta$, [\ion{O}{3}]$\lambda$5007, [\ion{O}{3}]$\lambda$4959, H$\alpha$, [\ion{N}{2}]$\lambda$6549, and [\ion{N}{2}]$\lambda$6585 were each modeled using a single Gaussian, with their velocities and velocity offsets tied together to ensure consistency. The broad components of H$\beta$ and H$\alpha$ were modeled using two Gaussians to account for the peak and extended wings. The best-fitting models were determined through $\chi^2$ minimization. From the resulting models, we extracted emission line fluxes and line widths. The continuum luminosities at 5100{\AA}, measured within a 40{\AA} window centered on the line 5100{\AA} ($\pm$20{\AA} on either side).
\begin{table*}
    \centering
    \caption{Variability statistics.}
    \label{tab:variability}
    \renewcommand{\arraystretch}{1}
    \setlength{\tabcolsep}{8pt}
    \begin{tabular}{c c c c c}
        \hline
        Source & Parameter & $g$-band & H$\beta$ & H$\alpha$ \\
        (1)&(2)&(3)&(4)&(5)\\
        \midrule
        Mrk 1048 & $F_{\mathrm{var}} (\%)$  & $7.30 \pm 0.47$ & $10.30 \pm 0.17$ & $6.75 \pm 0.14$ \\
        &$R_{\mathrm{max}}$& $1.42 \pm 0.03$ & $1.50 \pm 0.04$ & $1.27 \pm 0.03$ \\
        &Median & $14.19 \pm 0.01$ & $4.15 \pm 0.12 $& $13.6 \pm 2.50$\\
        Mrk 618 & $F_{\mathrm{var}} (\%)$& $4.20 \pm 0.24$ & $7.68 \pm 0.83$ & $13.91 \pm 2.00$ \\
        & $R_{\mathrm{max}}$ & $1.25 \pm 0.03$ & $1.42 \pm 0.06$ & $1.65 \pm 0.01$ \\
        & Median& $14.37 \pm 0.05$ & $2.50 \pm 0.34$ & $4.12 \pm 0.14$ \\
        \bottomrule
    \end{tabular}
    \begin{flushleft}
        \textbf{Note:} Variability statistics ($F_{\mathrm{var}}$), the maximum-to-minimum flux ratio ($R_{\mathrm{max}}$), and median values with their respective errors for $g$-band, H$\beta$, and H$\alpha$ in Mrk 1048 and Mrk 618. The median flux for the $g$-band is in magnitude, for H$\beta$ and H$\alpha$ emission lines, it is in $10^{-13}$ erg s$^{-1}$ cm$^{-2}$ \AA$^{-1}$.
    \end{flushleft}
\end{table*}
For TANSPEC spectra, the continuum was modeled using a power-law fit over selected line-free regions: [6502, 6645], [10736, 11005], [12700, 13000], [14765, 15755], and [19017, 19524]{\AA}. After subtracting the fitted continuum, spectral decomposition was performed in the H$\alpha$ region, using a single-Gaussian profile fit within the wavelength range 6400–6670{\AA}. The integrated flux of the H$\alpha$ line was then calculated by integrating the best-fitting Gaussian model.
\subsection{Light curve and Variability}
\label{sec:light curve}
For data obtained from ADFOSC and HFOSC, the emission line fluxes for H$\beta$ and H$\alpha$ were measured by integrating the area under the broad components modeled using best-fitting Gaussians from PyQSOFit, centered at 4861{\AA} and 6564{\AA}. The rest-frame wavelength ranges used for integration were [4780, 4940]{\AA} for H$\beta$ and [6450, 6680]{\AA} for H$\alpha$, with these regions consistently applied to both sources. However, the line widths and flux strengths differed significantly between Mrk 1048 and Mrk 618, reflecting source-specific kinematics and variability. In addition to the H$\alpha$ fluxes derived from TANSPEC (using single-Gaussian fitting), we employed {\tt PyCALI} once again to intercalibrate the H$\alpha$ light curves obtained with ADFOSC and HFOSC against those from TANSPEC. This step was essential to correct for noticeable flux offsets present in the TANSPEC spectra. However, since the TANSPEC dataset contains only a limited number of observations (5-6 data points), the use of {\tt PyCALI} introduces relatively larger uncertainties, which in turn affect the final flux calibration. To mitigate this issue, we smoothed the H$\alpha$ light curve by applying a five-point running average. The final intercalibrated H$\alpha$ and H$\beta$ emission line light curves were then used for lag measurements for the photometric $g$-band continuum light curve. Understanding short-term intrinsic variability is crucial for determining accurate lags. The final light curves are shown in Fig.~\ref{fig:lag plots}. The upper panel displays the $g$-band photometric continuum, with marked data points from different telescopes. The middle panel presents the H$\beta$ emission line light curve from ADFOSC and HFOSC observations, while the bottom panel shows the intercalibrated H$\alpha$ emission line light curve, including data points from TANSPEC. For Mrk 618, we have added the H$\beta$ flux points from Season 4 of \citet{MAHA2024} in our obtained H$\beta$ light curve for better cadence and correlation analysis.

Typical light curve parameters, including the fractional variability amplitude ($F_{\mathrm{var}}$), the maximum-to-minimum flux ratio ($R_{\mathrm{max}}$), and the median flux, are summarized in Table~\ref{tab:variability}. The fractional variability amplitude was calculated using the following equation \citep{RodriguezPascual1997}:
\begin{align} 
F_{\mathrm{var}} = \dfrac{\sqrt{\sigma^{2} - \langle \sigma^{2}_{\mathrm{err}} \rangle}}{\langle f \rangle} 
\end{align}
where $\sigma^{2}$ is the variance of the light curve, $\langle \sigma^{2}_{\mathrm{err}} \rangle$ is the mean square measurement uncertainty, and $\langle f \rangle$ is the mean flux.

\textit{Mrk 1048}: The properties of the light curves for Mrk 1048 are listed in the first three rows of Table~\ref{tab:variability}. The $F_{\mathrm{var}}$ for the photometric $g$-band continuum is approximately 7.30\%. The variability amplitudes for the H$\beta$ and H$\alpha$ emission lines are higher, at 10.3\% and 6.75\%, respectively. As expected, the median flux of the H$\alpha$ line exceeds that of H$\beta$, reflecting the intrinsic line strength differences between them \citep{Netzer2013, OsterbrockFerland2006}. The typical H$\alpha$/H$\beta$ flux ratio is around 3, and our results are broadly consistent with this value. Median flux values are reported in magnitudes for the $g$-band and in units of 10$^{-13}$ erg$\,$s$^{-1}\,$cm$^{-2}\,$\AA$^{-1}$ for the emission lines. The maximum-to-minimum flux ratios ($R_{\mathrm{max}}$) are 1.42, 1.50, and 1.27 for the $g$-band, H$\beta$, and H$\alpha$, respectively.

\textit{Mrk 618}: The light curve parameters for Mrk 618 are presented in the last three rows of Table~\ref{tab:variability}. The $g$-band continuum exhibits a relatively lower $F_{\mathrm{var}}$ of approximately 4.20\%. The emission lines show moderately higher variability, with $F_{\mathrm{var}}$ values of 7.68\% for H$\beta$ and 13.91\% for H$\alpha$. The greater scatter may partly influence the higher variability amplitude observed for H$\alpha$ in the TANSPEC data points and host galaxy dilution. Corresponding $R_{\mathrm{max}}$ and median flux values are also listed. As evident from Fig.~\ref{fig:lag plots}, the emission line light curves for Mrk 618 also show more pronounced variability than the photometric continuum. Additionally, the H$\alpha$ line strength is $\approx$3 times more prominent than H$\beta$, similar to Mrk 1048 median flux.
\section{Time lag measurement}
\label{sec:timelag analysis}
\subsection{ICCF and \textsc{JAVELIN}}
To measure the time between the continuum variations and the H$\beta$ and H$\alpha$ emission line responses, we employed two widely adopted techniques: the Interpolated Cross-Correlation Function (ICCF)\footnote{\url{https://bitbucket.org/cgrier/python_ccf_code/src/master/}} \citep{Gaskell1986, Peterson_1998} and the model-based code \textsc{JAVELIN}\footnote{\url{https://github.com/nye17/JAVELIN}} \citep{Zu2011, Zu2013}. Both methods have been extensively validated in the context of RM studies \citep[see][]{Peterson_1998, Peterson_2004, Bentz2014ApJ...796....8B, Barth2015, Woo2024}, and typically produce broadly consistent results \citep{Edelson2019}. The ICCF approach involves computing the cross-correlation function between the continuum and emission line light curves to identify the degree of correlation and the corresponding time lag. We explored a lag search range from –20 to +100 days, guided by previously reported lags for both sources (typically within $\sim$30 days; \citealt{U2022, MAHA2024}) and our total monitoring baseline. Following the approach discussed by \citet{Woo2024}, where it is emphasized that too wide a lag window may introduce spurious secondary peaks or dilute the correlation strength due to noise and sparse sampling, we tested narrower windows around the expected lag range. We found that the posterior lag distributions consistently peaked at the same locations, though with slightly lower correlation coefficients ($r_{\mathrm{max}}$) when using wider windows. When we refined the lag search window to -10 to +50 days for Mrk 1048 and Mrk 618, the primary ICCF peak remained prominent and $r_{\mathrm{max}}$ increased, suggesting improved sensitivity and reduced contamination from false correlations. Moreover, we do not have to deal with seasonal gaps as our monitoring period is roughly 5 months. To account for irregular time sampling, the ICCF method interpolates one light curve while holding the other fixed. Then, it averages the results of both configurations to construct the final cross-correlation function.

\begin{figure*}[t]
\centering
\includegraphics[height=9cm, width=18cm]{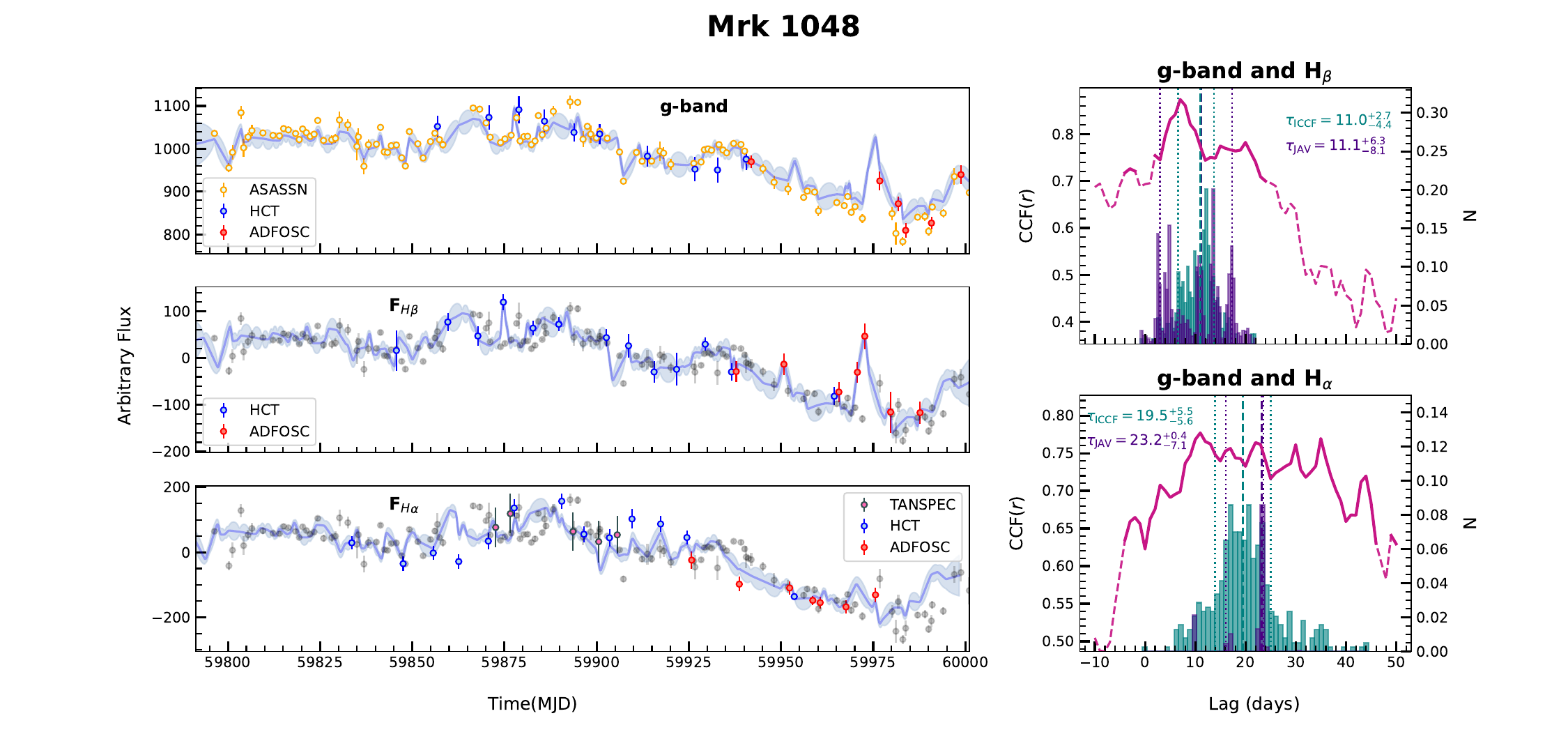}
\includegraphics[height=9cm, width=18cm]{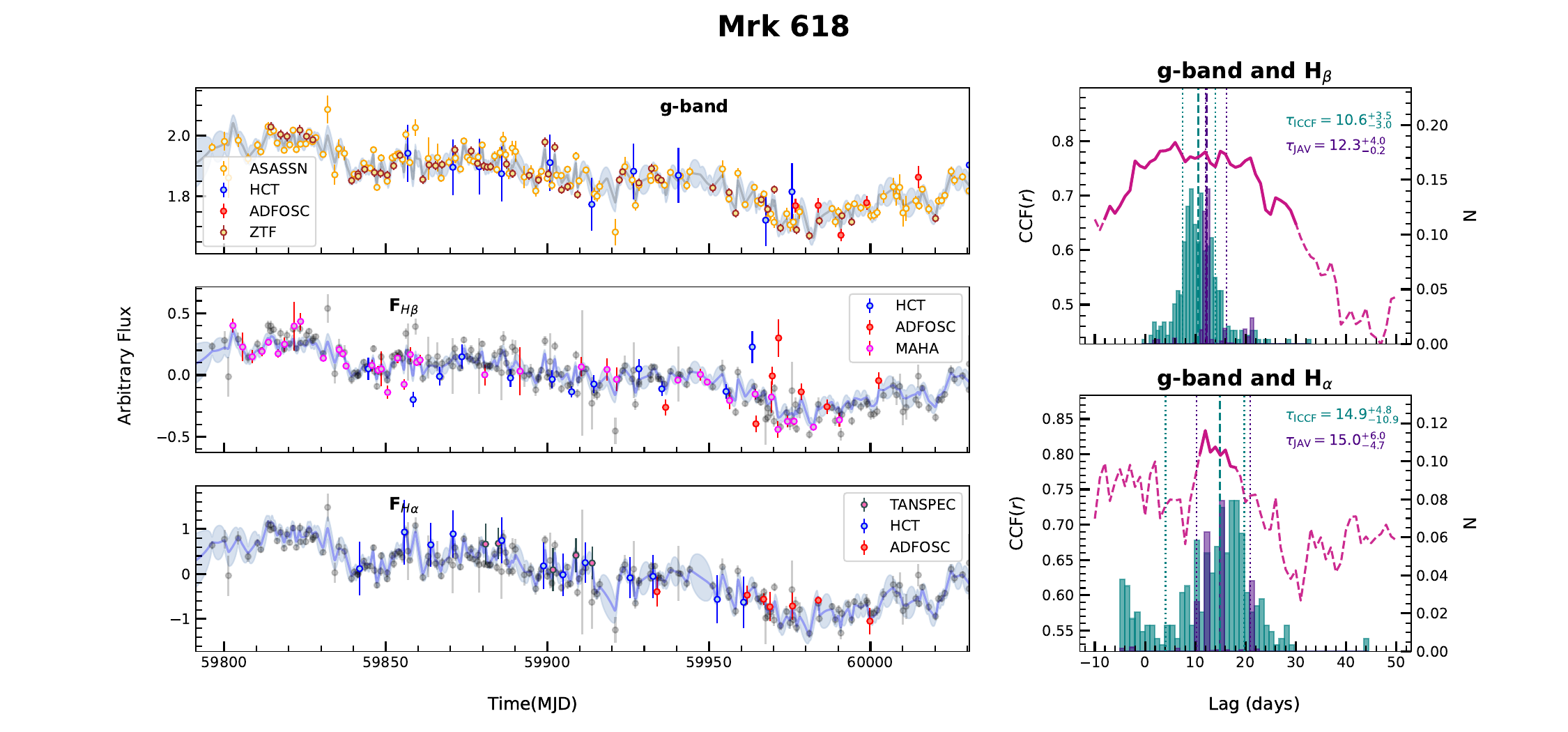}
\caption{Light curve plots for Mrk 1048 and Mrk 618. The upper panel shows a photometric $g$-band continuum with labelled data points from each telescope. The middle left and lower left panels display H$\beta$ and H$\alpha$ emission line fluxes in arbitrary units, with $g$-band continuum overlaid. These are mean-subtracted light curves and are matched by normalizing the $g$-band continuum light curve and shifting the emission line light curves to the final adopted lag values mentioned in Table \ref{tab:Time lag measurement}. The JAVELIN modelling for each light curve is shown in steel blue. For the H$\alpha$ light curve, we have smoothed it with five consecutive points using the running average method. The right upper and lower panels show the lag histograms from ICCF (teal) and JAVELIN (violet). These plots display the CCF $r_{\mathrm{value}}$ value on the left (pink) and the probability density(N) of the histograms on the right. The darker pink region of the $r_{\mathrm{value}}$ curve depicts 80\% of the centroid peak that is used to calculate the final ICCF lag. The dashed lines indicate the lags with 16th and 84th percentiles of the lag probability density.}
\label{fig:lag plots}
\end{figure*}

We used the flux randomization/random subset sampling (FR/RSS) Monte Carlo technique \citep{Peterson_1998, Peterson_2004} to quantify the uncertainty in lag measurements. This involves generating multiple realizations of the light curves by resampling and perturbing the data and computing the centroid lag ($\tau_{\mathrm{cent}}$) from the portion of the CCF above 80\% of the peak r$\mathrm{max}$, also highlighted in Fig. \ref{fig:lag plots}. The median of the resulting $\tau_{\mathrm{cent}}$ distribution is adopted as the best-measure lag. The derived lag values for both methods are summarised in Table~\ref{tab:Time lag measurement}.

\textsc{JAVELIN}, developed by \citet{Zu2011, Zu2013}, models AGN continuum variability using a damped random walk \citep[DRW;][]{Kelly2009, Kelly2014} process and derives emission line light curves by convolving the modeled continuum with a transfer function, typically a top-hat function. Uncertainties on the lag and other parameters are measured using a Markov Chain Monte Carlo (MCMC) approach, which provides statistical confidence intervals for the best-fit values. \textsc{JAVELIN} simultaneously models both continuum and emission line light curves. The DRW model has been shown to reproduce AGN variability on both short and long timescales across multiple bands, with some exceptions \citep[e.g.,][]{Mchardy2006, Mushotzky2011}. Compared to ICCF, \textsc{JAVELIN} often produces tighter constraints on lag measures \citep[e.g.,][]{Edelson2019, Yu2020}. Table~\ref{tab:Time lag measurement} presents the lag results.
\begin{table*}
\centering
\caption{Measured time delays (lags) for Mrk 1048 and Mrk 618 using ICCF, \textsc{JAVELIN}, and PyI$^{2}$CCF methods.}
\label{tab:Time lag measurement}
\setlength{\tabcolsep}{8pt}
\begin{tabular}{@{} l l cc c cc @{}}
\hline \hline
\multirow{2}{*}{\textbf{Source}} & \multirow{2}{*}{\textbf{Light curve}} &
\multicolumn{2}{c}{\textbf{ICCF}} &
\multicolumn{1}{c}{\textbf{JAVELIN}} &
\multicolumn{2}{c}{\textbf{PyI$^{2}$CCF}} \\
\cmidrule(lr){3-4}\cmidrule(lr){5-5}\cmidrule(lr){6-7}
& & Lag (days) & $r_{\mathrm{max}}$ & Lag (days)  & Lag (days)  & p-value\\
(1) & (2) & (3) & (4) & (5) & (6) & (7) \\
\midrule
\multirow{2}{*}{Mrk~1048}
  & $g$-band vs H$\beta$ & \(11.0^{+2.7}_{-4.4}\) & 0.9 & \(11.1^{+6.3}_{-8.1}\) &  \(11.4^{+3.1}_{-4.1}\) & 0.09 \\
  & $g$-band vs H$\alpha$& \(19.5^{+5.5}_{-5.6}\) & 0.7 & \(23.2^{+0.4}_{-7.1}\) &  \(20.1^{+6.7}_{-6.1}\) & 0.16 \\

\multirow{2}{*}{Mrk~618}
  & $g$-band vs H$\beta$ & \(10.6^{+3.5}_{-3.0}\) & 0.7 & \(12.3^{+4.0}_{-0.2}\) &  \(12.0^{+7.5}_{-7.0}\) & 0.06 \\
  & $g$-band vs H$\alpha$& \(14.9^{+4.8}_{-10.9}\) & 0.8 & \(15.0^{+6.0}_{-4.7}\) &  \(14.0^{+4.4}_{-8.8}\) & 0.04 \\
\bottomrule
\end{tabular}

\begin{tablenotes}[flushleft]
\footnotesize
\item \textbf{Note.} Lags are in the observer frame for g-band vs H$beta$ and H$\alpha$ emission line light curves. Columns: (1) Source; (2) light-curve chosen to calculate the lag; (3) ICCF centroid lag; (4) cross-correlation coefficient $r_{\mathrm{max}}$; (5) JAVELIN lag; (6) PyI\(^{2}\)CCF lag; (7) PyI$^{\mathrm{2}}$CCF Null hypothesis value (p). The lag search range for both sources, Mrk 1048 and Mrk 618, is between -10 to 50 days.
\end{tablenotes}

\end{table*}

\textit{Mrk 1048:} The ICCF analysis yields a lag of $11.0^{+2.7}_{-4.4}$ days with a well-defined peak and a maximum cross-correlation coefficient ($r_{\mathrm{max}}$) of 0.88, whereas \textsc{JAVELIN} yields $11.1^{+6.3}_{-8.1}$ days between the $g$-band and H$\beta$, being consistent within the errors. For the $g$-band vs. H$\alpha$, both ICCF and \textsc{JAVELIN} give lag values of $19.5^{+5.5}_{-5.6}$ and $23.2^{+0.4}_{-7.1}$ days, respectively, with $r_{\mathrm{max}} > 0.75$ for the ICCF. The H$\beta$ lags were comparatively shorter than the H$\alpha$ lag values from each method, hinting at the BLR stratification \citep{Bentz2010}. The lag distributions are visualised in Fig.~\ref{fig:lag plots} (right panels). For $g$-band vs. H$\beta$, the ICCF and \textsc{JAVELIN} lag distributions are confined within 0 to 25 days with a very sharp peak near measured lag values. In contrast, for $g$-band vs. H$\alpha$, \textsc{JAVELIN} posterior distribution is narrower than the ICCF lag distribution with three sharp peaks, which is widely spread till 25 days. The lag results for H$\beta$ are broadly consistent with those reported by \citet{U2022} as part of the LAMP 2016 campaign, where a rest-frame lag of $\tau_{\mathrm{cent}} = 9.0^{+9.4}_{-7.4}$ days was measured with a $r_{\mathrm{max}}$ shaped more likely as a flat top. Our monitoring over a longer duration yields a more sharply defined lag with a higher $r_{\mathrm{max}} = 0.88$, improving upon the earlier constraints.

\textit{Mrk 618:} The lag measurement from ICCF and \textsc{JAVELIN} measures of $10.6^{+3.5}_{-3.0}$ and $12.3^{+4.0}_{-0.2}$ days, respectively, with a maximum correlation coefficient ($r_{\mathrm{max}}$) of approximately 0.72, as reported in Table \ref{tab:Time lag measurement}. The measured lag for the $g$-band vs. H$\alpha$ light curves is consistent and similar to H$\beta$ within error, with $14.9^{+4.8}_{-10.9}$ days ($r_{\mathrm{max}}$=0.81) obtained via ICCF and  $15.0^{+6.0}_{-4.7}$ days via \textsc{JAVELIN}. The lag distributions for these pairs, shown in the upper and lower right panels of the second row in Fig.~\ref{fig:lag plots}, reveal stronger confined peaks for H$\beta$ and H$\alpha$. Uncertainties for both methods were derived from the entire probability of the lag distributions. The lag results are very similar to those reported by \citet[][hereafter referred to as the Monitoring AGNs with H$\beta$ Asymmetry (MAHA) survey]{MAHA2024}. They provided a range of lag values obtained over four seasons, and our results for H$\beta$ are in agreement with its lag value measured in the fourth season.

\subsection{Simulations}
\label{sec:simulations}
To assess the robustness of our measured lags and determine whether the observed time sampling is adequate for reliable lag detection, we performed extensive light curve simulations. For each source (Mrk 1048 and Mrk 618), we generated mock continuum light curves based on the Damped Random Walk (DRW) model, with parameters tuned to match the variability amplitude and timescale of the observed $g$-band continuum. Importantly, the simulated light curves were constructed with the same temporal sampling and data gaps as the real observations, thereby preserving realistic observational conditions. The Emission line light curves for H$\beta$ and H$\alpha$ were then synthesized by shifting, smoothing, and scaling the mock continuum light curves using the observed lag values derived from our ICCF analysis (see Table~\ref{tab:Time lag measurement}). We applied both ICCF and \textsc{JAVELIN} time-series analysis methods to each realization in order to recover the input lag. This process was repeated for 1000 independent simulations for each emission line and each source. 

\begin{figure}
    \centering
    \includegraphics[width=1\linewidth]{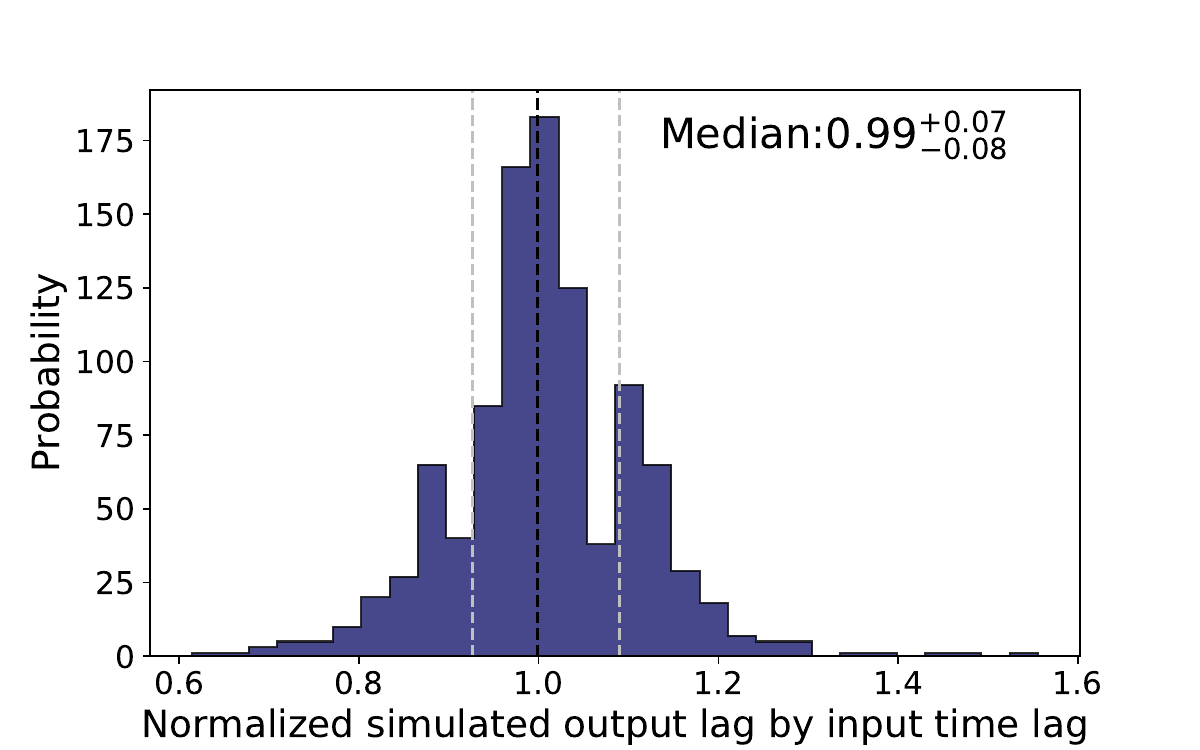}
    \caption{We simulated light curves for the $g$-band continuum and the H$\beta$ and H$\alpha$ emission lines with 1000 independent realizations each. The figure shows the probability distribution of the ratio between the recovered lag and the input ICCF lag (11 days) for the Mrk 1048 $g$-band versus H$\beta$ case with the ICCF method. The quoted median of the distribution is close to unity, indicating that the observed sampling and noise levels are sufficient to recover the intrinsic lag reliably.}
    \label{fig:lag_simulation}
\end{figure}

Fig.~\ref{fig:lag_simulation} shows the distribution of recovered-to-input lag ratios from the ICCF simulations. The distributions are strongly centred around unity, confirming that the input lags can be accurately recovered under the actual cadence and noise conditions. These results validate the significance and reliability of our measured lags for both H$\beta$ and H$\alpha$ in Mrk 1048 and Mrk 618.

We have also employed the publicly available PyI$^{2}$CCF code\footnote{\url{https://github.com/legolason/PyIICCF/}}, developed by \citet{Guo2022} and based on the method described in \citet{U2022}. This approach evaluates the statistical significance of the lag measurements and provides an independent check on the reliability of the ICCF method. The method is grounded on the null hypothesis that, when two uncorrelated random light curves are cross-correlated, the maximum correlation coefficient ($r_{\mathrm{max}}$) should be greater than or equal to the observed value $r_{\mathrm{max,obs}}$ obtained from the actual light curves. To test this, the code generates a large ensemble of mock light curves from a damped random walk (DRW) model with the same noise properties and cadence as the observed data \citep[see also][]{U2022, pandey_spectroscopic_2022, Buitrago2023, Woo2024}. In this work, we generated 5000 mock realizations of the continuum, H$\beta$, and H$\alpha$ light curves. The resulting lag measurements and their significance are summarized in Table~\ref{tab:Time lag measurement}. Alongside the ICCF and JAVELIN results, the PyI$^{2}$CCF lag values and the corresponding null-hypothesis $p$-values are reported. Following the reliability criteria $p \leq 0.2$ \citep{U2022, Guo2022, Woo2024} and $r_{\mathrm{max}} > 0.5$, we confirm that all lag measurements listed in Table~\ref{tab:Time lag measurement} are robust. Notably, $r_{\mathrm{max}}$ values remain in the range $0.7$–$0.9$, further supporting the reliability of the ICCF results.

\section{Effect of detrending}
\label{sec: detrend}
In our analysis, a linear trend is apparent in the light curves of Mrk~1048, while it is less pronounced in Mrk~618. To account for this, we applied a linear detrending procedure by fitting a straight line to both the continuum and emission-line light curves and subtracting the best-fit model from the original data. This approach isolates short-term intrinsic variability while minimizing the impact of long-term drifts \citep[see, e.g.,][]{Zhang2019,Woo2024}, without introducing artificial fluctuations that could arise from higher-order polynomial fits. The detrended light curves and corresponding lag measurements are shown in Fig.~\ref{fig:detrended result}. Time lags were re-estimated using both ICCF and \textsc{JAVELIN}, and the results are summarized in Table~\ref{tab:detrended_lags}. The posterior lag distributions displayed in Fig.~\ref{fig:detrended result} exhibit broader spreads or multiple peaks in some cases, reflecting weaker correlations. Consistently, the maximum cross-correlation coefficients ($r_{\mathrm{max}}$) decrease across all cases—dropping to $\leq 0.6$ for Mrk~1048 and below 0.3 for Mrk~618. To further evaluate the statistical significance of these lags, we employed PyI$^{2}$CCF, with the corresponding null-hypothesis $p$-values listed in Table~\ref{tab:detrended_lags}. For Mrk~1048, the $p$-values exceed 0.25, while for Mrk~618 they are even higher (up to 0.72 for the $g$-band vs.\ H$\alpha$ correlation), indicating that detrending substantially reduces the apparent lag significance. The detrended results are discussed in Sec.~\ref{sec:appendix}; however, due to their lower significance, these lags were not adopted as our final measurements.
\section{Black hole mass measurement}
\label{sec:Black hole mass measurement}
\subsection{Mean and RMS spectrum}
We constructed the mean spectrum and the root-mean-square (rms) spectrum using the following definitions:
\begin{align}
\bar{F(\lambda)} =\dfrac{1}{N}\sum_{i=0}^{N-1}F_{i}(\lambda),
\end{align} 
and
\begin{align}
S(\lambda) =\sqrt{\Big[\dfrac{1}{N-1}\sum_{i=1}^{N}(F_{i}(\lambda)- \bar{F(\lambda)}^{2})\Big]}
\end{align}
where $F_i(\lambda)$ represents the $i$-th spectrum in a set of $N = 25$ spectra collected during the monitoring campaign for each source.

\begin{figure*}[t]
    \centering
    \includegraphics[width=1\linewidth]{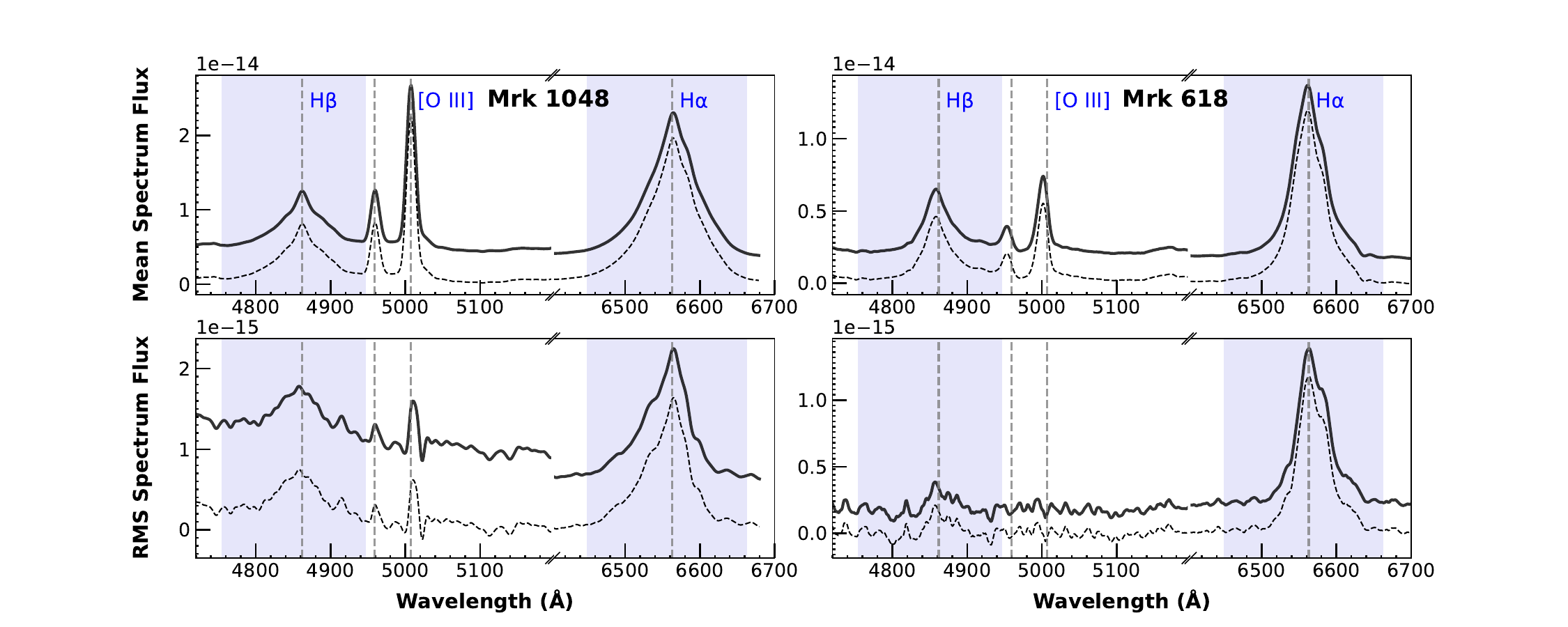}
    \caption{The mean and rms spectra for Mrk 1048 (left) and Mrk 618 (right). The solid indicates the original spectrum, while the dotted lines are the continuum-subtracted spectrum.}
    \label{fig:mean_rms_spectrum}
\end{figure*}

The mean spectrum ($\bar{F}(\lambda)$) represents the average flux at each wavelength across all epochs and typically exhibits a high signal-to-noise ratio (S/N). In contrast, the rms spectrum ($S(\lambda)$) characterises the variability at each wavelength, highlighting regions with significant temporal flux changes. For the final rms spectrum, we subtracted the contribution from measurement noise by estimating the observed variance at each wavelength and removing the average noise variance. This was carried out with inverse-variance weighting across epochs to account for differing uncertainties. To quantify the reliability of the rms spectrum, bootstrap resampling was used to estimate the $1\sigma$ uncertainties, providing robust confidence intervals on the intrinsic variability. Fig.~\ref{fig:mean_rms_spectrum} shows the mean and rms spectra for Mrk 1048 and Mrk 618 in the left and right panels, respectively. In each panel, solid and dotted lines denote the spectra before and after subtraction of the power-law continuum. Prominent emission line regions are visible in both panels. The mean spectrum, benefiting from higher S/N, reveals strong features including both narrow and broad emission lines. The rms spectrum, while having lower S/N, also displays narrow emission lines that ideally should be absent. This is likely due to the use of data from different instruments and telescopes, where factors such as spectral alignment, instrumental resolution, and seeing conditions during observations can significantly impact the construction of the rms spectrum. Consequently, as shown in Fig.~\ref{fig:mean_rms_spectrum}, both sources exhibit [\ion{O}{3}] emission lines, which are more prominent in Mrk 618 than in Mrk 1048. The rms spectrum of Mrk 1048 closely resembles that presented in \citet{U2022}. Although noisier, the rms spectrum effectively isolates variable components by suppressing contributions from non-varying features such as narrow emission lines and host galaxy starlight. This makes it a valuable diagnostic for identifying intrinsically variable broad-line components. However, due to its lower signal-to-noise and sensitivity to noise fluctuations, measuring emission line widths from the rms spectrum remains challenging.  
\subsection{Line width and black hole mass measurement}
To measure the black hole masses, we measured the full width at half maximum (FWHM) and the line dispersion ($\sigma_{\mathrm{line}}$) of the H$\beta$ and H$\alpha$ emission line from both the mean and rms spectra after continuum subtraction. The FWHM was calculated by identifying the wavelengths corresponding to 50\% of the maximum flux on the blue and red sides of the emission line profile, denoted as $\lambda_l$ and $\lambda_r$, respectively. The FWHM is then obtained as the difference $\lambda_r - \lambda_l$ \citep{Peterson_2004}. To compute the line dispersion, we first determined the flux-weighted centroid of the line using the expression.
\begin{align} 
\lambda_{0} = \dfrac{\int \lambda f_{\lambda} d\lambda}{\int f_{\lambda} d\lambda}
\end{align}
followed by calculating the second moment of the profile as
\begin{align} 
\sigma_{\mathrm{line}}^{2} = \dfrac{\int \lambda^{2} f_{\lambda} d\lambda}{\int f_{\lambda} d\lambda} - \lambda_{0}^{2}.
\label{eq:sigma}
\end{align}
Assuming that the motion of the BLR gas is dominated by the gravitational potential of the central black hole, we measured the black hole mass using the virial relation from Eq.\ref{eq:virial eqn}. We adopted a scaling factor $f = 4.47$ for line dispersion-based measurements, and $f = 1.12$ for those based on FWHM \citep{Woo2015}. To estimate the uncertainties in the emission line widths and, consequently, the black hole mass, we employed a Monte Carlo bootstrap method following the approach of \citet{Peterson_2004}. For each realization, $N$ spectra were randomly selected with replacement from the original set of $N$ nightly spectra, from which the mean and rms spectra were reconstructed. Similarly, for the continuum regions adjacent to each emission line, we incorporated their flux uncertainties into the analysis. At every Monte Carlo realization, the continuum level was randomly varied within its measured error range and then subtracted from the emission-line region. This step is important because the exact placement of the continuum directly affects the shape and strength of the residual emission line. By explicitly including this uncertainty, we ensure that errors arising from imperfect continuum determination are consistently carried into the final estimates of the emission-line width. Consequently, the derived values of FWHM and $\sigma_{\mathrm{line}}$ reflect not only the random noise present in the spectra but also the systematic uncertainty associated with continuum subtraction, providing a more realistic and reliable error budget. For H$\beta$ and H$\alpha$, the continuum sidebands were randomly varied within $\pm 10$\,\AA\ \citep{Barth2015} of the nominal windows ([4780,4940]\,\AA\ for H$\beta$ and [6450,6680]\,\AA\ for H$\alpha$). A total of 5000 realizations were generated, each yielding a perturbed line profile from which both FWHM and $\sigma_{\mathrm{line}}$ were measured. The median of the resulting distributions was adopted as the final value of the line width, and the 16th and 84th percentiles defined the 1$\sigma$ confidence interval. The instrumental resolution was also subtracted from the obtained FWHM and $\sigma_{\mathrm{line}}$. The uncertainty in black hole mass measurement is measured by propagating the errors of lag $\tau \pm \sigma_{\tau}$ and line width $\Delta V \pm \sigma_{\Delta V}$. These values were consistently used to measure the black hole masses from the mean and rms spectra.
We used the lag values derived from the ICCF method to measure the black hole masses for both sources. Table~\ref{tab:bh_mass_bounds} presents the results of black hole mass measurements based on both the H$\alpha$ and H$\beta$ emission lines, using line widths obtained from the mean and rms spectra. The line widths are resolution-corrected with respect to ADFOSC and HFOSC instrumental resolution. 

In the case of \textit{Mrk 1048}, black hole mass measures for H$\beta$ emission line are between $4.81^{+1.5}_{-1.6}\times 10^{7} M_{\odot}$ to $5.71^{+1.7}_{-1.4}\times 10^{7} M_{\odot}$ considering FWHM based on the mean and rms spectrum, respectively. Additionally, using the $\sigma_{\mathrm{line}}$ defined from Eq. \ref{eq:sigma} as a line width estimator, the black hole mass is in $4.23^{+1.3}_{-1.4}\times 10^{7} M_{\odot}$ to $6.30^{+2.0}_{-2.1}\times 10^{7} M_{\odot}$. For H$\alpha$ emission line, the black hole mass with FWHM is in the range of $4.54^{+1.3}_{-1.3}\times 10^{7} M_{\odot}$ to $5.42^{+1.5}_{-1.6}\times 10^{7} M_{\odot}$ and $4.81^{+1.3}_{-1.4}\times 10^{7} M_{\odot}$ to $5.40^{+1.5}_{-1.5}\times 10^{7} M_{\odot}$ with $\sigma_{\mathrm{line}}$. Notably, the FWHM and $\sigma_{\mathrm{line}}$ of the H$\beta$ line are broader than those of H$\alpha$, consistent with previous findings that H$\beta$ tends to trace higher velocity gas in the BLR \citep[e.g.,][]{greene_estimating_2005, WangandShen2019}. This trend becomes more evident in the rms spectrum. The broader line widths seen in the rms spectra compared to the mean spectra may result from the reduced signal-to-noise ratio at the line wings in the mean spectrum, which can lead to underestimation of the true line width.

For \textit{Mrk 618}, the emission line widths are generally narrower than those of Mrk 1048. From the mean spectrum, for the H$\beta$ emission line, the black hole mass ranges from $1.15^{+0.4}_{-0.3}\times 10^{7} M_{\odot}$ and $1.75^{+0.6}_{-0.5} \times 10^{7} M_{\odot}$ using FWHM and $\sigma_{\mathrm{line}}$, respectively. For the H$\alpha$ emission line, the black hole mass for Mrk 618 is measured to be $1.36^{+0.4}_{-0.7}\times 10^{7} M_{\odot}$ and $2.15^{+0.6}_{-0.9}\times 10^{7} M_{\odot}$, respectively. Similarly, for the rms spectrum, the masses are $1.06^{+0.4}_{-0.3}\times 10^{7} M_{\odot}$ and $0.62^{+0.2}_{-0.2}\times 10^{7} M_{\odot}$ for H$\beta$, whereas for H$\alpha$, the masses are $1.67^{+0.5}_{-0.8}\times 10^{7} M_{\odot}$ and $1.19^{+0.4}_{-0.6}\times 10^{7} M_{\odot}$, respectively using FWHM and $\sigma_{\mathrm{line}}$.

It is important to note that single-epoch black hole mass estimates are sensitive to the choice of line width used in the virial equation. While FWHM is commonly adopted in single-epoch mass measurements, RM studies involving multiple emission lines have shown that $\sigma_{\mathrm{line}}$ offers a more reliable estimator of the virial velocity \citep{Peterson_2004}. Since the rms spectrum effectively isolates variable components by removing non-varying features such as narrow emission lines and host galaxy contributions, it is generally more robust for black hole mass estimation. Therefore, we adopt the $\sigma_{\mathrm{line}}$ measurements from the rms spectra as our preferred measures of black hole mass. Hence, the black hole mass of Mrk 1048 is 6.30$^{+2.0}_{-2.1} \times 10^{7} M_{\odot}$ as measured using both the H$\beta$ and 4.81$^{+1.3}_{-1.4}\times 10^{7} M_{\odot}$ for H$\alpha$ emission lines, whereas for Mrk 618 the mass is calculated as 6.2$^{+2.0}_{-2.0}\times 10^{6} M_{\odot}$ and 1.19$^{+0.4}_{-0.6}\times 10^{7} M_{\odot}$ using H$\beta$ and H$\alpha$ emission, respectively.
\begin{table*}
\centering
\caption{H$\beta$ and H$\alpha$ Line Widths and Black Hole Mass Measurements with Lower and Upper Limits.}
\label{tab:bh_mass_bounds}
\renewcommand{\arraystretch}{1.3}
\begin{tabular}{lcccccc}
\toprule
Source & Line & FWHM(km$\,$ s$^{-1}$) & $M_\mathrm{BH}$ ($\times 10^7 M_{\odot}$) & $\sigma_{\mathrm{line}}$ (km$\,$ s$^{-1}$) & $M_\mathrm{BH}$ ($\times 10^7 M_{\odot}$) \\
(1)&(2)&(3)&(4)&(5)&(6)\\
\midrule
\multicolumn{6}{c}{\textbf{Mean Spectrum Measurements}} \\
\midrule
Mrk 1048  & H$\beta$ & $4677^{+276}_{-268}$  & $4.81^{+1.5}_{-1.6}$  & $2193^{+45}_{-47}$  & $4.23^{+1.3}_{-1.4}$ \\
          & H$\alpha$ & $3643 ^{+186}_{-174}$  & $5.42^{+1.5}_{-1.6}$  & $1817 ^{+27}_{-26}$  & $5.40^{+1.5}_{-1.5}$ \\
Mrk 618   & H$\beta$ & $2261 ^{+128}_{-139}$ & $1.15^{+0.4}_{-0.3}$  & $1499^{+49}_{-55}$  & $1.75^{+0.6}_{-0.5}$ \\
          & H$\alpha$ & $2113 ^{+134}_{-83}$ & $1.36^{+0.4}_{-0.7}$  & $1327 ^{+21}_{-20}$  & $2.15^{+0.6}_{-0.9}$ \\
\midrule
\multicolumn{6}{c}{\textbf{RMS Spectrum Measurements}} \\
\midrule
Mrk 1048  & H$\beta$ & $4975^{+260}_{-251}$  & $5.71^{+1.7}_{-1.4}$  & $2678^{+71}_{-77}$  & $6.30^{+2.0}_{-2.1}$ \\
          & H$\alpha$ & $3333^{+211}_{-234}$ & $4.54^{+1.3}_{-1.3}$  & $1716^{+63}_{-61}$  & $4.81^{+1.3}_{-1.4}$ \\
Mrk 618   & H$\beta$ & $2172^{+139}_{-136}$  & $1.06^{+0.4}_{-0.3}$  & $832^{+44}_{-69}$  & $0.62^{+0.2}_{-0.2}$\\
          & H$\alpha$ & $2335^{+113}_{-99}$  & $1.67^{+0.5}_{-0.8}$  & $987^{+61}_{-63}$  & $1.19^{+0.4}_{-0.6}$\\
\bottomrule
\end{tabular}
\vspace{2mm}
\begin{flushleft}
\small \textbf{Note:} Columns are (1) Object name, (2) the line used for calculations, (3) the Full-width at half-maximum of the emission line in km$\,$ s$^{-1}$. (4) Black hole mass measured using FWHM. (5) $\sigma_\mathrm{line}$ (km$\,$ s$^{-1}$): Line dispersion (second moment) of the emission line profile. (6) Black hole mass measured using $\sigma_\mathrm{line}$. Unit of black hole masses is $\times 10^7 M_{\odot}$.
\end{flushleft}
\end{table*}
\section{Discussion}
\label{sec:Discussion}
\subsection{Size–Luminosity Relation}
\label{sec:R_L}
\textit{Mrk 1048} was previously monitored as part of the LAMP 2016 campaign \citep{U2022} measuring H$\beta$ rest-frame time lag of $\tau_{\mathrm{cent}} = 9.0^{+9.4}_{-7.4}$ days and $r_{\mathrm{max}} = 0.6$. In comparison, our monitoring spanning October 2022 to March 2023 yielded a rest-frame H$\beta$ lag of $10.5^{+2.6}_{-4.2}$ days and a higher $r_{\mathrm{max}} = 0.9$. In a more recent effort, \citet{Sobrino2025} included Mrk 1048 (NGC 985) in a large-scale photometric RM (PRM) campaign using narrow-band targeting the H$\alpha$ emission line in nearby Seyfert galaxies ($0.015 < z < 0.05$). For Mrk 1048, they obtained a single-epoch spectrum and modeled the broad H$\alpha$ emission line. Using their refined PRM formalism, they derived a rest-frame H$\alpha$ time lag of $21.3 \pm 0.7$ days, which is comparable with our measured rest frame lag of $18.7^{+5.3}_{-5.4}$ days.

\textit{Mrk 618} was previously observed in a 2012 RM campaign by \citet{DeRosa2018}, where no significant H$\beta$ lag was detected due to a shorter monitoring period. In contrast, the recent multi-year campaign by \citet{MAHA2024} reported lag detections across four seasons (2019–2023), with H$\beta$ lags ranging from $9.2^{+1.6}_{-2.3}$ to $30.9^{+10.6}_{-7.2}$ days. The strongest signal occurred in Season 2, whereas the lag of Season 3 was deemed less reliable due to a dual-peaked cross-correlation function. Our current RM campaign independently confirms a strong reverberation signature in Mrk 618. We detect rest frame time lag of $10.2^{+3.4}_{-2.9}$ days (ICCF) for $g$-band vs H$\beta$ and $14.4^{+4.6}_{-10.5}$ days for $g$-band vs H$\alpha$ (Table~\ref{tab:Time lag measurement}), which is consistent with the best lag value of 15.2$^{+2.4}_{-2.3}$ reported by MAHA campaign in their Season 4 Observations. These results reaffirm the presence of a responsive BLR in Mrk 618.

\begin{figure}
    \centering
    \includegraphics[height=9cm, width=9cm]{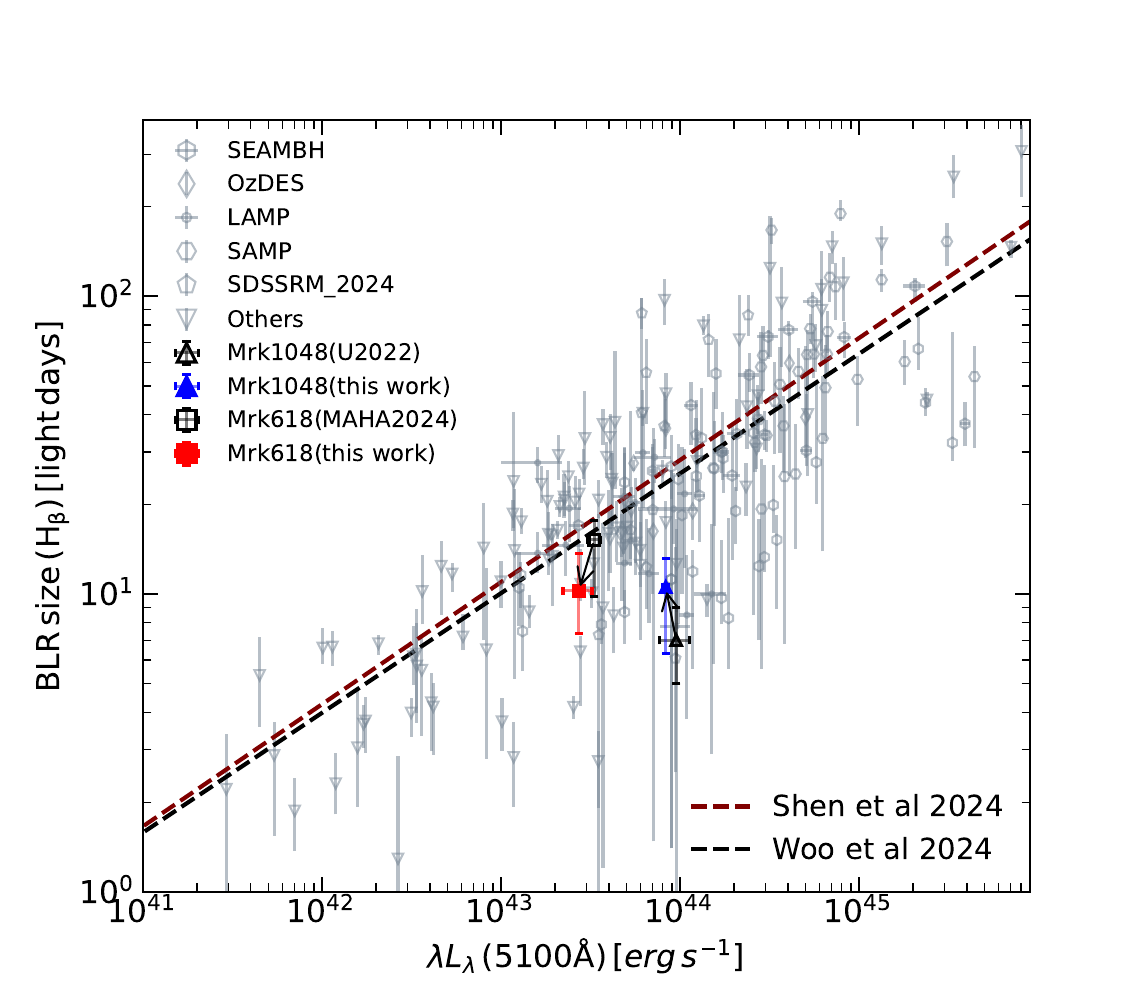}
    \caption{The plot illustrates the relationship between the H$\beta$ BLR size and the optical continuum luminosity at 5100{\AA}. A range of reverberation-mapped (RM) AGN samples from the literature are overplotted for comparison. These include SEAMBHs \citep[][grey open circles]{Du_2016, Du_2018,Hu2021,Li2021}, SDSSRM-2024 sources \citep[][grey open squares]{Shen2024}, OzDES AGNs \citep[][grey open diamonds]{Malik2023}, LAMP sources \citep[][grey open circles]{U2022}, and SAMP sources \citep[][grey open hexagons]{Woo2024}. Additional RM sources from various studies \citep[][grey open inverted triangles]{bentz_low-luminosity_2013, Park_2017, Rakshit2019, Bonta2020, Rakshit2020A&A, pandey_spectroscopic_2022} are also included. The maroon and black dashed line represents the best-fit $R$–$L$ relation as reported by \citet{Woo2024} and \citet{Shen2024}, respectively. Our target sources, Mrk 1048 and Mrk 618, are marked with a filled blue triangle and a filled orange circle, respectively. Mrk 1048 moves towards upper left, while Mrk 618 moves towards lower right, to the best-fit relation. These are still lying closer to the best-fit relation.}
    \label{fig:R-Lhbeta}
\end{figure}

We placed Mrk 1048 and Mrk 618 on the H$\beta$-based BLR size–luminosity ($R_{\mathrm{BLR}}$–$L_{5100}$) plot using the empirical relation:
\begin{align} \label{eq:R_L}
\log \left( \frac{R_{\mathrm{BLR}}}{\text{lt-day}} \right) = K + \alpha \log \left( \frac{\lambda L_{\lambda}(5100\ \text{\AA})}{10^{44}\ \text{erg s}^{-1}} \right)
\end{align}
where the slope $\alpha = 0.41$ and intercept $K = 1.45$ with intrinsic scatter 0.32 dex, as calibrated by \citet[][S2024 hereafter]{Shen2024} and with slope $\alpha=0.402$ and intercept $K =1.405$ by \citet[][W2024 hereafter]{Woo2024} with intrinsic scatter 0.23 dex. For Mrk 1048, with $L_{5100} = 8.30 \times 10^{43}$ erg s$^{-1}$, the predicted H$\beta$ BLR sizes are $R_{\mathrm{BLR}} = 26.2$ light-days (S2024) and $23.6$ light-days (W2024), whereas our RM measurement yields a smaller lag of $10.5$ light-days. In contrast, for Mrk 618, with $L_{5100} = 2.71 \times 10^{43}$ erg s$^{-1}$, the predicted sizes are $16.5$ light-days (S2024) and $15.1$ light-days (W2024), while our measured lag is comparable at $10.2$ light-days. These discrepancies are within 1$\sigma$ limit from the global $R_{\mathrm{BLR}}$–$L_{5100}$ relation for Mrk 1048. This is visualised in Fig.~\ref{fig:R-Lhbeta}.

Correcting for host galaxy contamination is essential for accurately determining AGN luminosities, as it significantly impacts the size–luminosity relation and can introduce substantial uncertainties if unaccounted for. We applied the empirical host-fraction relation from \citet{Jalan2023}, which is based on the host contamination measurement from SDSS spectra, and estimated host contributions of $\sim43.7\%$ for Mrk 1048 and $\sim57.4\%$ for Mrk 618. This yields host-subtracted AGN continuum luminosities of $L_{5100,\mathrm{AGN}} \approx 4.67 \times 10^{43}$erg s$^{-1}$ and $\approx 1.15 \times 10^{43}$ erg s$^{-1}$, respectively. Recalculating the expected BLR sizes with these corrected values, we obtain $R_{\mathrm{BLR}} = 20.6$ and $18.8$\,light-days (S2024 and W2024, respectively) for Mrk 1048, and $11.6$ and $10.7$\,light-days for Mrk 618. 

\begin{figure}
    \centering
    \includegraphics[height=9cm, width=9cm]{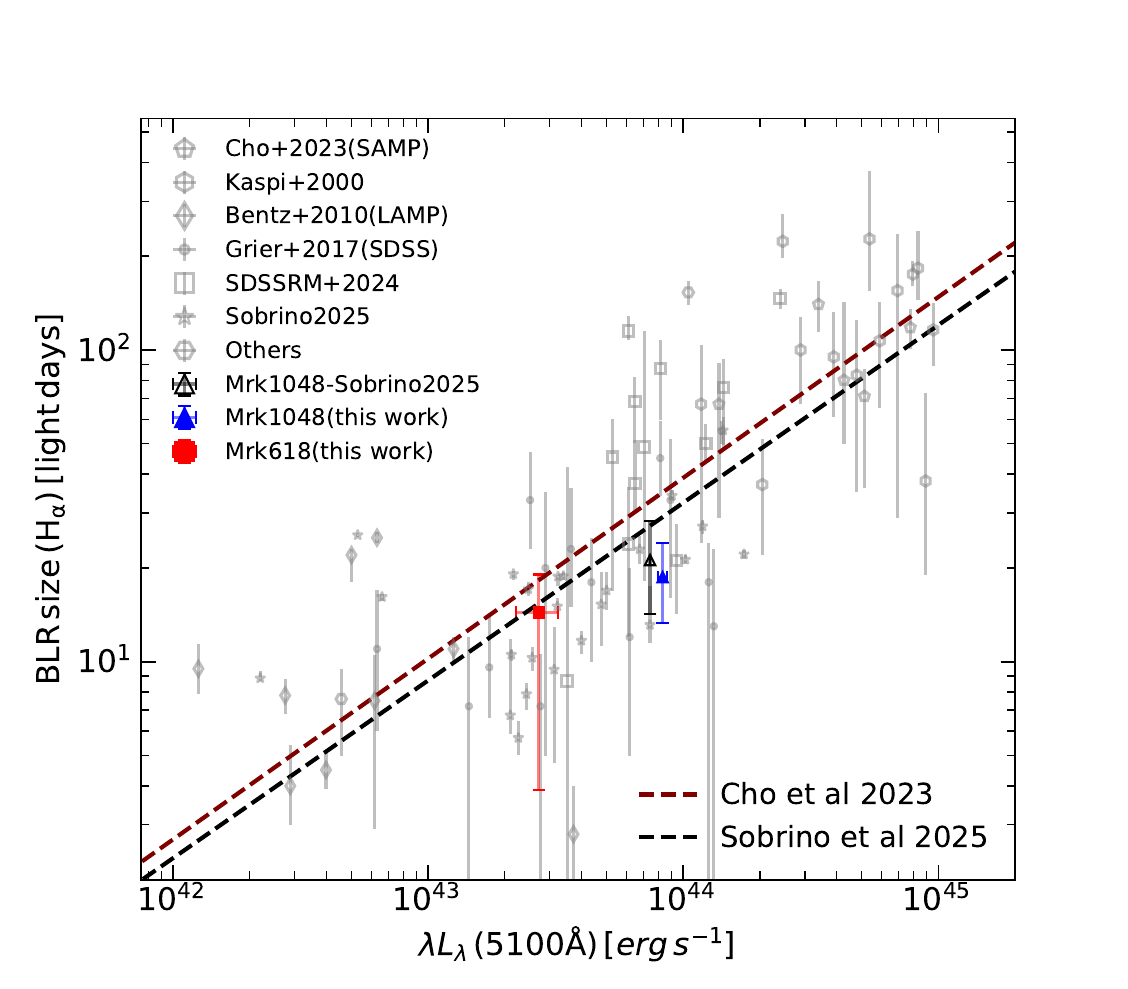}
    \caption{The relationship between the H$\alpha$ BLR size and the monochromatic continuum luminosity at 5100{\AA} is shown. Reverberation-mapped (RM) sources with H$\alpha$ lag measurements from previous studies are overplotted for comparison, including those from \citet{Kaspi2000}, \citet{Bentz2010}, \citet{Grier2017}, \citet{Cho_2020}, and \citet{Woo2024}, \citet{Sobrino2025}, \citet{Shen2024}, along with additional sources from \citet{Sergeev2017}, \citet{Feng_2021}, and \citet{Li_2022}. The maroon and black dashed line represents the best-fit $R$–$L$ relation as derived by \citet{Cho2023} and \citet{Sobrino2025}, respectively. Our target AGNs, Mrk 1048 and Mrk 618, are plotted as a filled blue triangle and a filled orange circle, respectively. The previous measure of H$\alpha$ for Mrk 1048 \citep{Sobrino2025} is also marked with an open triangle. Mrk 618 lies closer to the H$\alpha$ best-fit relation, while Mrk 1048 is showing a slightly more offset than \citet{Sobrino2025} study.}
    \label{fig:R-Lhalpha}
\end{figure}

To cross-check our spectro-photometric H$\alpha$ lag measurements, we estimated the BLR sizes using the directly measured 5100\,\AA\ continuum luminosities and the H$\alpha$-based $R_{\mathrm{BLR}}$–$L_{5100}$ relations from \citet[][C2023, hereafter]{Cho2023} and \citet[][S2025, hereafter]{Sobrino2025}. For this, we used Eq. \ref{eq:R_L} for the H$\alpha$ emission line with $K = 1.51$, $\alpha = 0.57$, and scatter =0.32 dex based on S2025, and $K = 1.59$, $\alpha = 0.58$, with scatter =0.31$\,$dex based on C2023. Using the total luminosities, the predicted BLR sizes for Mrk 1048 are $34.8$ and $29.1$ light-days (C2023 and S2025, respectively), compared to our measured lag of $18.7$ light-days. For Mrk 618, the predictions are $18.2$ and $15.4$\,light-days, while our measured lag is comparable at $14.4$ light-days. Applying host-galaxy correction, the AGN-only 5100$\,$\AA\ luminosities yield revised $R_{\mathrm{BLR}}$ values of $25.1$ and $21.0$\,light-days (C2023 and S2025, respectively) for Mrk 1048, and $11.2$ and $9.5$\,light-days for Mrk 618. The Mrk 1048 is within a factor of $\sim$1.2, and Mrk 618 shows a factor of $\sim$1.4 times larger measured lag than predicted from the H$\alpha$-based $R_{\mathrm{BLR}}$–$L_{5100}$ relation. This is shown in Fig.~\ref{fig:R-Lhalpha}, consistent with the deviation observed in the H$\beta$-based scaling. 

The standard $R_{\mathrm{BLR}}$–$L_{5100}$ relation tends to overpredict the BLR sizes of high-accretion AGNs with strong Fe\,\textsc{ii} emission. \citet{Du_2019} proposed a refined relation incorporating the relative Fe\,\textsc{ii} strength, $\mathrm{R}_{\mathrm{FeII}} = \mathrm{EW}_{\mathrm{FeII}} / \mathrm{EW}_{\mathrm{H\beta}}$, showing that higher $\mathrm{R}_{\mathrm{FeII}}$ values correspond to shorter H$\beta$ lags at fixed luminosity. For Mrk 618, we measure $\mathrm{R}_{\mathrm{FeII}} \approx 1.50$, suggesting a predicted BLR size below 7 light-days. However, our measured H$\beta$ lag is slightly longer at $\sim$10.2 light-days, indicating that Mrk 618 deviates minimally from this trend.
\subsection{Stratification of BLR}
We computed the ratio of H$\alpha$ to H$\beta$ lags to assess the ionisation stratification of the BLR in our sources. For Mrk 1048, this ratio is:
\begin{equation}
    \frac{\tau_{\mathrm{H\alpha}}}{\tau_{\mathrm{H\beta}}} = \frac{18.7^{+5.3}_{-5.4}}{10.5^{+2.6}_{-4.2}} \approx 1.7^{+0.6}_{-0.9},
\end{equation}
and for Mrk 618:
\begin{equation}
    \frac{\tau_{\mathrm{H\alpha}}}{\tau_{\mathrm{H\beta}}} = \frac{14.4^{+4.6}_{-10.5}}{10.2^{+3.4}_{-2.9}} \approx 1.4^{+0.6}_{-1.1}.
\end{equation}
These ratios indicate that in Mrk 1048, the H$\alpha$-emitting region lies farther out in the BLR than the H$\beta$-emitting region, consistent with expectations from photoionisation stratification. In contrast, the ratio for Mrk 618 suggests a more co-spatial origin of the two lines, potentially linked to a flatter radial ionisation profile or more compact BLR geometry.

Our result for Mrk 1048 notably contrasts with the value of $\tau_{\mathrm{H\alpha}}/\tau_{\mathrm{H\beta}} = 2.9^{+1.4}_{-1.1}$ reported by S2025, who adopted the H$\beta$ lag of $\tau_{\mathrm{H\beta}} = 7.4^{+9.7}_{-9.4}$ days from \citet{U2022} and derived a rest-frame H$\alpha$ lag of $21.3 \pm 0.7$ days. Their reported ratio was among the highest in their sample and interpreted as strong evidence for radial stratification in the BLR. In contrast, our updated measurements yield a more moderate ratio of $\tau_{\mathrm{H\alpha}}/\tau_{\mathrm{H\beta}} \approx 1.7^{+0.6}_{-0.9}$, which lies closer to the average and median values reported in the literature. Specifically, S2025 reported a mean H$\alpha$/H$\beta$ lag ratio of $1.6 \pm 0.8$ and a median of across their full sample, consistent with earlier studies by \citet{Kaspi2000}, \citet{Bentz2010}, and \citet{Shen2024}, which reported average ratios around 1.4. Our revised result thus falls well within this statistically expected range, suggesting that S2025 high-ratio estimates may have been inflated due to sparse cadence, low signal-to-noise, or systematics in non-uniform spectral sampling. This reinforces the value of dedicated, well-calibrated, and high-cadence monitoring campaigns in accurately tracing BLR stratification.

Furthermore, our measured ratio for Mrk 1048 is more consistent with the typical range of 1.2-1.8 found by C2023 for high-luminosity AGNs using a recalibrated H$\alpha$-H$\beta$ BLR structure analysis. This comparison underscores the importance of uniform and simultaneous spectral monitoring for reliably interpreting BLR stratification and dynamics.
\subsection{Black hole masses}
\label{sec:Previous literature}
\textit{Mrk 1048:} Line width comparisons show that while \citet{U2022} found FWHM$_{\mathrm{mean}} =$ 4830 $\pm$ 80 km$\,$s$^{-1}$ and $\sigma_{\mathrm{mean}} =$ 1840 $\pm$ 58 $\,$km$\,$s$^{-1}$, our results are FWHM$_{\mathrm{mean}} = $4677$^{+276}_{-268} \,$km$\,$s$^{-1}$ and $\sigma_{\mathrm{mean}} = $2193$^{+45}_{-47}\,$km$\,$s$^{-1}$. However, our broader rms spectrum values (FWHM$_{\mathrm{rms}} = $4975$^{+260}_{-251} \,$km$\,$s$^{-1}$, $\sigma_{\mathrm{rms}} =$ 2678$^{+71}_{-77}\,$km$\,$s$^{-1}$) are compared to those from the LAMP campaign (4042 $\pm$ 406 $\,$km$\,$s$^{-1}$ and 1726 $\pm$ 76$\,$km$\,$s$^{-1}$) indicate that our data captured a larger portion of the line variability, possibly tracing higher-velocity components of the BLR. Consequently, the black hole mass measured by \citet{U2022} using the rms $\sigma_{\mathrm{line}}$ was $2.2 \times 10^7\,M_\odot$, which is significantly lower than our result of $6.30 \times 10^7\,M_\odot$, a factor of $\sim$3 difference. Aditionally, our directly measured continuum luminosity at 5100{\AA} is $8.30 \pm 0.35 \times 10^{43}$ erg s$^{-1}$, is consistent with the $9.5 \pm 1.8 \times 10^{43}$ erg s$^{-1}$ reported by \citet{U2022}. The difference in the black hole mass estimate could be due to a lack of host galaxy correction, variability differences in both monitoring campaigns, and broader line width measurement. While \citet{U2022} identified infalling BLR kinematics using velocity-resolved RM, we were not able to perform such an analysis due to the limited number of epochs in our campaign. \citet{Sobrino2025} reported a black hole mass of $M_{\mathrm{BH}} = 9.12^{+0.30}_{-0.34} \times 10^7\,M_\odot$ using the FWHM of the H$\alpha$ line. Our mass measurement is in closer agreement with theirs.  Their host-subtracted continuum luminosity at 5100{\AA} was derived using the flux variation gradient (FVG) method, while the Eddington ratio of $0.109$ (which is closer to our estimate of $0.094$) was computed via Fe II emission line strength. The slight discrepancy in mass measurement may stem from differences in the method used for BLR size measurement (their Photometric RM compared to our spectroscopic RM), emission line width treatment, single-epoch assumptions, or line modeling details.

For \textit{Mrk 618}, the integrated line widths derived from our spectra yield FWHM values of 2261$^{+128}_{-139}\,$ km $\,$s$^{-1}$ (mean) and $2172^{+139}_{-136}\,$ km$\,$ s$^{-1}$ (rms) for H$\beta$, and corresponding $\sigma_{\mathrm{line}}$ values of $1499^{+49}_{-55}\,$ km$\,$ s$^{-1}$ and $832^{+44}_{-69}\,$ km$\,$ s$^{-1}$. These values are slighlty smaller with the range of FWHM = 2387–3219 km$\,$ s$^{-1}$ and $\sigma_{\mathrm{line}}$ = 1279-1650 km$\,$s$^{-1}$ reported in MAHA. Additionally, our H$\alpha$ measurements display consistent broadening behavior, indicating a stable BLR geometry across multiple lines and epochs. Furthermore, our decomposition of the H$\beta$ profile reveals a mild asymmetry that leans toward the blue wing, which aligns with the findings of the MAHA campaign, that show evolving line asymmetry across seasons. Specifically, MAHA reported H$\beta$ asymmetry values ranging from $-0.256$ to $-0.144$, suggesting a shift from disk-like dynamics (Season 2) to an outflow-dominated geometry (Season 4). While we do not perform velocity-resolved lag measurements in our current dataset, the presence of asymmetry in the line profiles hints at similar kinematic complexities. Our directly measured continuum luminosity at 5100{\AA} is $2.72 \pm 0.50 \times 10^{43}$ erg s$^{-1}$, which reduces to $1.15 \times 10^{43}$ erg s$^{-1}$ after correcting for the host galaxy contribution, which is lower than the MAHA Season 4 value of $3.31 \pm 0.28 \times 10^{43}$ erg s$^{-1}$. The decreasing trend in L$_{5100}$ across the MAHA seasons from 4.73 to 3.31 $\times 10^{43}$ erg s$^{-1}$ is aligned with the lower luminosity recorded in our campaign, further supporting the observed decline in AGN activity.
\subsection{Implication of SARM observation}
\label{sec:SARM observation}
To estimate the angular extent of the BLR, we used our directly measured BLR radii from H$\beta$ and H$\alpha$ time lags, along with angular diameter distances derived under a standard $\Lambda$CDM cosmology ($H_0 = 70\ \mathrm{km\ s^{-1}\ Mpc^{-1}}$, $\Omega_M = 0.3$, $\Omega_\Lambda = 0.7$). The angular size was calculated using the relation:
\begin{align}
\xi_{\mathrm{BLR}} = \frac{R_{\mathrm{BLR}}}{D_A},
\end{align}
where $R_{\mathrm{BLR}}$ is the BLR radius in parsecs, and $D_A$ is the angular diameter distance in megaparsecs. The resulting angular size was then converted to microarcseconds ($\mu\mathrm{as}$). 

For Mrk 1048, our measured H$\beta$ and H$\alpha$ lags of $10.5^{+2.6}_{-4.2}$ and $18.7^{+5.3}_{-5.4}$ light-days corresponds to BLR radii of $0.0085^{+0.0020}_{-0.0034}$ pc and $0.0152^{+0.0043}_{-0.0044}$ pc, given that $1\ \mathrm{lt\text{-}day} \approx 0.0008$ pc, adopting a redshift of $z = 0.043$. With an angular diameter distance of $D_A = 168.0$ Mpc, the angular sizes are $\xi_{\mathrm{BLR}} \approx 10.1^{+2.4}_{-4.0}\,\mu\text{as}$ (H$\beta$) and $15.5^{+4.4}_{-4.5}\,\mu\text{as}$ (H$\alpha$). For Mrk 618 ($z = 0.034$), our H$\beta$ and H$\alpha$ lags of $10.2^{+3.4}_{-2.9}$ and $14.4^{+4.6}_{-10.5}$ light-days yield BLR radii of $0.0083^{+0.0027}_{-0.0024}$ pc and $0.0117^{+0.0038}_{-0.0085}$ pc, respectively. With $D_A = 137.1$ Mpc, the corresponding angular sizes are $\xi_{\mathrm{BLR}} \approx 12.6^{+4.1}_{-3.5}\,\mu\text{as}$ (H$\beta$) and $17.1^{+5.4}_{-10.1}\,\mu\text{as}$ (H$\alpha$).

These angular sizes are generally smaller than those predicted in \citet{Wang2020}, who report, for instance, $\xi_{\mathrm{BLR}} = 46.4 \mu\mathrm{as}$ for Mrk 1048 based on an assumed BLR size of 48.6 light-days. The discrepancy likely reflects differences in BLR size measurements across epochs and methods. Nonetheless, the scales we derive remain within the reach of the interferometric resolution of GRAVITY. 

The Pa$\alpha$ emission line ($\lambda_{\mathrm{rest}} = 1.875\, \mu$m), commonly used in spectroastrometry, is redshifted to $1.956\, \mu$m for Mrk 1048, which lies around the window edge within GRAVITY's $K$-band coverage ($1.95$–$2.45\, \mu$m). However, for Mrk 618, the redshifted Pa$\alpha$ line appears at $1.94\, \mu$m, just outside the lower limit of this band, limiting its accessibility. In such cases, alternative broad emission lines such as Brackett~$\gamma$ (Br$\gamma$, $\lambda_{\mathrm{rest}} = 2.17\, \mu$m), He\,\textsc{i} ($2.06\, \mu$m), H$_2$ 1--0 S(1) ($2.12\, \mu$m), and [Si\,\textsc{vi}] ($1.96\, \mu$m) are viable options. For both Mrk 1048 and Mrk 618, these lines redshift to within the $K$-band window and offer promising alternatives, however challenging for spatially resolving the BLR kinematics using GRAVITY/GRAVITY+.

To improve SARM measurements, efforts should be made to achieve better consistency and reduce systematic uncertainties on the data \citep[see][]{Wang2020}. A key improvement would be the use of the same broad emission line, such as Pa$\alpha$ or H$\beta$, in both spectroastrometry and reverberation mapping observations, ensuring that both methods probe the same physical region of the broad-line region (BLR). Although one of the goals of our campaign was to observe the Infrared lines, the number of such epochs is too few to perform a detailed investigation for the lag measurement. Additionally, conducting these observations (RM and SA) jointly or within the dynamical timescale of BLR can minimise biases arising from temporal variations in the BLR structure. Incorporating information from velocity-resolved reverberation mapping and polarimetry can help constrain key physical parameters such as inclination, opening angle, and the degree of ordered motion in the BLR. Expanding the sample size to include well-monitored AGNs and improving the precision of interferometric phase measurements will further reduce both statistical and systematic errors.
\section{Conclusion}
\label{sec:conclusion}
We observed the sources Mrk 1048 and Mrk 618 among the seven sources selected from \citet{Wang2020} with the aim of performing SARM studies. Our spectro-photometric weekly cadence monitoring of Mrk 1048 and Mrk 618 was conducted between October 2022 and March 2023 using Optical and NIR instruments such as ADFOSC and TANSPEC mounted at the 3.6-m DOT and the HFOSC mounted on the 2-m HCT. Broad-band photometric monitoring was done using V, R, and SDSS r filters immediately before spectroscopy. Intercalibrated $g$-band light curves, using ASAS-SN and ZTF data, were adopted as the primary continuum driver after alignment via PyCALI. This work covers the first part of the SARM, i.e., RM analysis. We gathered the following results for Mrk 1048 and Mrk 618.
\begin{enumerate}
    \item Mrk 1048 shows moderate continuum variability with a fractional variability amplitude ($F_{\mathrm{var}}$) of 7.30\% in the $g$-band, while the H$\beta$ and H$\alpha$ emission lines exhibit slightly higher variability, at 10.30\% and 6.75\%, respectively. The corresponding maximum-to-minimum flux ratios ($R_{\mathrm{max}}$) are 1.42 ($g$-band), 1.50 (H$\beta$), and 1.27 (H$\alpha$). In contrast, Mrk 618 displays a lower continuum variability ($F_{\mathrm{var}} \sim 4.20\%$) but a more pronounced variability in its emission lines, with 7.68\% for H$\beta$ and a notably higher 13.91\% for H$\alpha$. The emission lines in both sources exhibit greater flux variability than the continuum, consistent with reverberation expectations, and H$\alpha$ consistently appears stronger and more variable than H$\beta$ in both objects.
    \item Using ICCF-based lags and optical luminosities, both sources were placed on the $R_{\mathrm{BLR}}$–$L_{5100}$ relation for H$\beta$ and H$\alpha$ emission lines. For Mrk 1048, the H$\beta$ lag is 10.5$^{+2.6}_{-4.2}$ days with $L_{5100} = 8.30 \pm 0.35 \times 10^{43}$ erg s$^{-1}$, while H$\alpha$ gives a longer lag of 18.7$^{+5.3}_{-5.4}$ days. For Mrk 618, the H$\beta$ and H$\alpha$ lags are 10.2$^{+3.4}_{-2.9}$ and 14.4$^{+4.6}_{-10.5}$ days, respectively, with a lower luminosity of $L_{5100} = 2.71 \pm 0.50 \times 10^{43}$ erg s$^{-1}$. Both sources exhibit mildly deviated and broadly consistent with the $R_{\mathrm{BLR}}$–$L_{5100}$ relation, with Mrk 618 appearing slightly offset in H$\beta$, however, more in H$\alpha$. Whereas, for Mrk 1048 deviated in H$\beta$ and more closer to H$\alpha$, reflecting structural or ionization differences in their BLRs.
    \item Black hole mass measures for Mrk 1048 range from 4.2 to $6.3 \times 10^{7}\,M_{\odot}$, with H$\beta$. For Mrk 618, narrower line widths yield lower masses, ranging from $0.6$ to $1.7 \times 10^{7}\,M_{\odot}$ depending on the choice of line-width and emission line. As $\sigma_{\mathrm{line}}$ from rms spectra best isolates the variable BLR component, we adopt these as our preferred values: $6.30^{+2.0}_{-2.1} \times 10^{7}\,M_{\odot}$ for Mrk 1048 and 1.19$^{+0.4}_{-0.6} \times 10^{7}\,M_{\odot}$ for Mrk 618.    
\end{enumerate}
\begin{acknowledgments}
SR acknowledges partial support from the Science and Engineering Research Board (SERB), Department of Science and Technology (DST), Government of India, through the Start-up Research Grant (SRG) no. SRG/2021/001334. J.H.W. acknowledges the support from the Basic Science Research Program through the National Research Foundation of the Korean Government (grant No. NRF-2021R1A2C3008486). A.K.M. acknowledges the support from the European Research Council (ERC) under the European Union’s Horizon 2020 research and innovation program (grant No. 951549).
This study makes use of data obtained from the 3.6-m Devasthal Optical Telescope (DOT), a National Facility operated by the Aryabhatta Research Institute of Observational Sciences (ARIES), and the 2-m Himalayan Chandra Telescope (HCT). ARIES is an autonomous institute under the DST, Government of India. We are grateful to the scientific and technical staff at ARIES for their invaluable support during DOT observations. The TIFR–ARIES Near Infrared Spectrometer (TANSPEC) mounted on DOT was developed jointly by TIFR, ARIES, and MKIR, Hawaii. We thank the 3.6-m DOT and IR astronomy teams at TIFR for their assistance during TANSPEC observations. We also acknowledge the support provided by the staff at the Indian Astronomical Observatory (IAO), Hanle, and the CREST facility in Hosakote, both operated by the Indian Institute of Astrophysics (IIA), Bengaluru, for enabling observations with HCT.
\end{acknowledgments}
\vspace{5mm}
\facilities{3.6-m DOT(ADFOSC, TANSPEC), 2-m HCT(HFOSC)}
\software{{\tt mapspec} \citep{MAPSPEC}, PyQSOFit \citep{GuoPyqsofit, guo_legolasonpyqsofit_2023}, {\tt PyCALI} \citep{Pycali2014}, PyCCF \citep{Peterson_1998}, \textsc{JAVELIN} \citep{Zu2011, Zu2013}, PyI$^{2}$CCF  \citep{Guo2022}}
\clearpage

\appendix
\restartappendixnumbering
\section{Analysis of detrended lightcurve}
\label{sec:appendix}
\begin{figure*}[t]
\centering
\includegraphics[height=9cm, width=18cm]{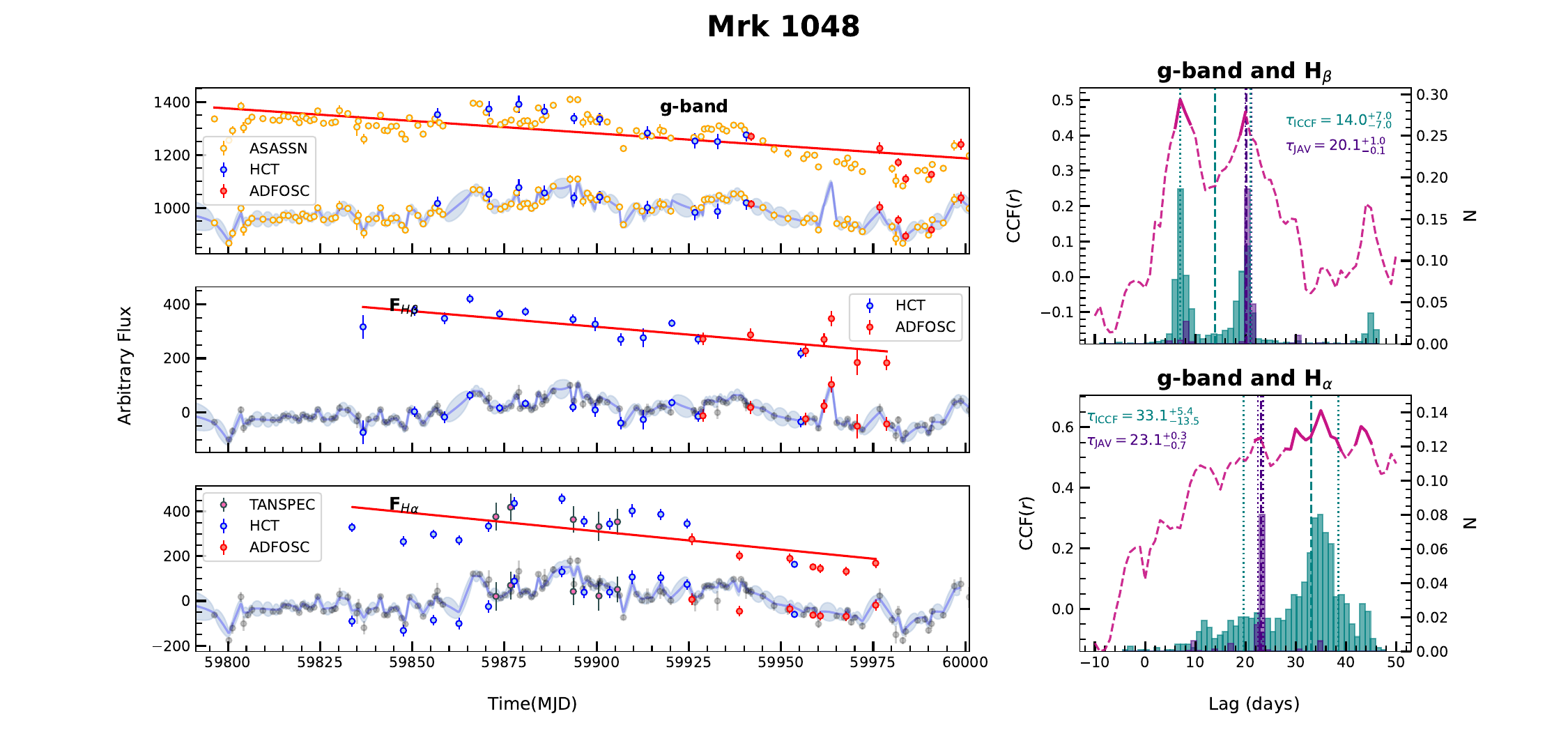}
\includegraphics[height=9cm, width=18cm]{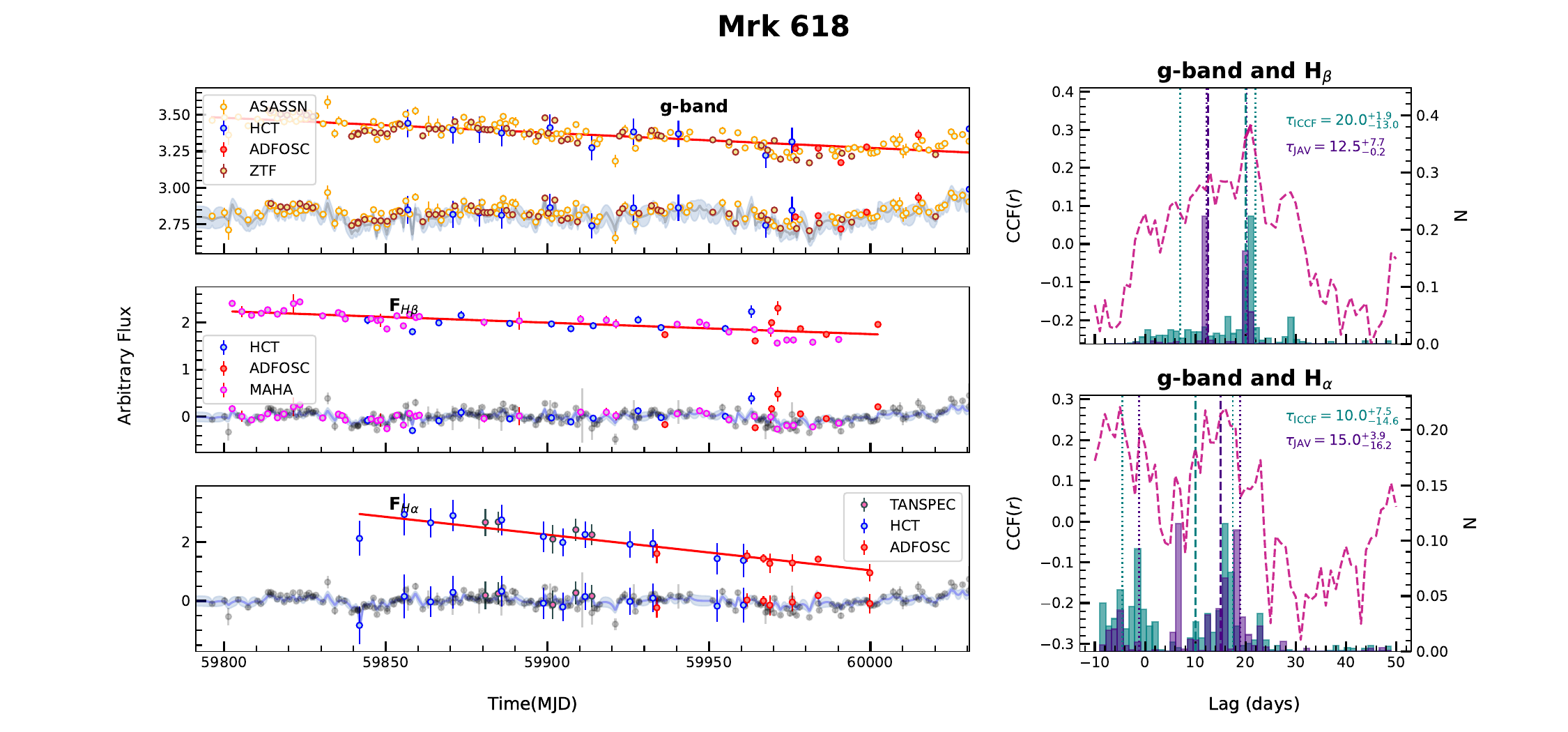}
\caption{Detrended light curve analysis for Mrk~1048 and Mrk 618. The top-left panel shows the photometric $g$-band continuum with data points from different telescopes labeled. The middle and bottom left panels display the H$\beta$ and H$\alpha$ emission-line flux variations (in arbitrary units), each fitted with a best-fit linear trend shown in red, that has been subtracted to obtain the detrended light curves. The $g$-band continuum is overlaid for comparison, normalized and shifted to the final adopted lag values listed in Table~\ref{tab:detrended_lags}. The \textsc{JAVELIN} model for each light curve is shown in steel blue. The right panels show the corresponding lag distributions from ICCF (teal histograms) and \textsc{JAVELIN} (violet histograms). The magenta dashed curve represents the cross-correlation function (CCF), with the left axis showing $r_{\mathrm{max}}$ and the right axis showing the probability density $N$. The darker magenta region marks the central 80\% of the CCF peak used to determine the ICCF centroid lag. The vertical dashed lines indicate the 16th and 84th percentile uncertainties of the lag distributions.}
\label{fig:detrended result}
\end{figure*}

We detrended the continuum and emission-line light curves by fitting a straight line to each and subtracting the corresponding best-fit model to obtain the residual (detrended) variations as described in Sec. \ref{sec: detrend}. Fig. ~\ref{fig:detrended result} and Table \ref{tab:detrended_lags} illustrates, the light curves and lag results for the entire campaign after detrending.
\begin{table*}
\centering
\caption{Detrended time delays (lags) for Mrk 1048 and Mrk 618 using ICCF, \textsc{JAVELIN}, and PyI$^{2}$CCF.}
\label{tab:detrended_lags}
\setlength{\tabcolsep}{8pt}
\begin{tabular}{@{} l l cc c cc @{}}
\hline \hline
\multirow{2}{*}{\textbf{Source}} & \multirow{2}{*}{\textbf{Light curve}} &
\multicolumn{2}{c}{\textbf{ICCF}} &
\multicolumn{1}{c}{\textbf{JAVELIN}} &
\multicolumn{2}{c}{\textbf{PyI$^{2}$CCF}} \\
\cmidrule(lr){3-4}\cmidrule(lr){5-5}\cmidrule(lr){6-7}
& & Lag (days) & $r_{\mathrm{max}}$ & Lag (days)  & Lag (days)  & p-value\\
(1) & (2) & (3) & (4) & (5) & (6) & (7) \\
\midrule
\multirow{2}{*}{Mrk~1048}
  & $g$-band vs H$\beta$ & \(14.0^{+7.0}_{-7.0}\) & 0.50 & \(20.1^{+1.0}_{-0.1}\) & \(16.58^{+5.0}_{-9.1}\) & 0.30 \\
  & $g$-band vs H$\alpha$& \(33.1^{+5.4}_{-13.5}\) & 0.62 & \(23.1^{+0.3}_{-0.7}\) & \(33.05^{+6.5}_{-11.5}\) & 0.25 \\

\multirow{2}{*}{Mrk~618}
  & $g$-band vs H$\beta$ & \(20.0^{+1.9}_{-13.0}\) & 0.30 & \(12.5^{+7.7}_{-0.2}\) & \(20.47^{+1.0}_{-10.0}\) & 0.27 \\
  & $g$-band vs H$\alpha$& \(10.0^{+7.5}_{-14.6}\) & 0.28 & \(15.0^{+3.9}_{-16.2}\) & \(15.0^{+3.0}_{-18.5}\) & 0.72 \\
\bottomrule
\end{tabular}

\begin{tablenotes}[flushleft]
\footnotesize
\item \textbf{Note.} All lags here are from \emph{detrended} light curves and are quoted in the observer frame. Columns: (1) Source; (2) light-curve pair; (3) ICCF centroid lag; (4) cross-correlation coefficient $r_{\mathrm{max}}$; (5) \textsc{JAVELIN} lag; (6) PyI$^{2}$CCF lag; (7) PyI$^{2}$CCF null-hypothesis $p$-value.
\end{tablenotes}
\end{table*}

\bibliographystyle{aasjournalv7}

\end{document}